\documentclass[aps,pre,twocolumn,floatfix]{revtex4-2}
\pdfoutput=1 
\usepackage{graphicx}
\usepackage{graphics}
\usepackage{latexsym}
\usepackage{amsmath, amsthm, amssymb, wasysym}
\usepackage{xcolor}
\usepackage{epsfig}
\usepackage{soul}
\usepackage{cancel}
\usepackage{hyperref}
\usepackage{orcidlink}
\usepackage{tikz}
\usetikzlibrary{shapes.geometric}

\definecolor{mplblue}{HTML}{1F77B4}
\definecolor{mplorange}{HTML}{FF7F0E}
\definecolor{mplgreen}{HTML}{2CA02C}
\definecolor{mplred}{HTML}{D62728}
\definecolor{rbcpurple}{HTML}{7B3294}
\definecolor{rbcteal}{HTML}{1A9E8F}
\definecolor{legendgray}{gray}{0.4}

\DeclareRobustCommand{\mksq}[1]{\tikz[baseline=-0.6ex]{%
  \fill[#1] (-0.7ex,-0.7ex) rectangle (0.7ex,0.7ex);}}
\DeclareRobustCommand{\mkhex}[1]{\tikz[baseline=-0.6ex]{%
  \fill[#1] (0,0.85ex) -- (0.736ex,0.425ex) -- (0.736ex,-0.425ex)
  -- (0,-0.85ex) -- (-0.736ex,-0.425ex) -- (-0.736ex,0.425ex) -- cycle;}}
\DeclareRobustCommand{\mkcirc}[1]{\tikz[baseline=-0.6ex]{%
  \fill[#1] (0,0) circle (0.75ex);}}
\DeclareRobustCommand{\mkdia}[1]{\tikz[baseline=-0.6ex]{%
  \fill[#1] (0,0.95ex) -- (0.95ex,0) -- (0,-0.95ex) -- (-0.95ex,0) -- cycle;}}
\DeclareRobustCommand{\mktri}[1]{\tikz[baseline=-0.6ex]{%
  \fill[#1] (0,0.9ex) -- (0.85ex,-0.6ex) -- (-0.85ex,-0.6ex) -- cycle;}}

\DeclareRobustCommand{\cubsym}[1]{\mksq{#1}}   % Figs. 1,2,4: simple cubic
\DeclareRobustCommand{\hexsym}[1]{\mkhex{#1}}  % Figs. 1,2,4: simple hexagonal
\DeclareRobustCommand{\rectsym}[1]{\mkcirc{#1}}% Fig. 3: rectangular (2D)
\DeclareRobustCommand{\orthsym}[1]{\mksq{#1}}  % Fig. 3: orthorhombic (3D)
\DeclareRobustCommand{\diasym}[1]{\mkdia{#1}}  % Fig. 5: square (2D)
\DeclareRobustCommand{\trisym}[1]{\mktri{#1}}  % Fig. 5: simple cubic (3D)

\DeclareRobustCommand{\legsimsq}{\mksq{legendgray}}
\DeclareRobustCommand{\legsimhex}{\mkhex{legendgray}}
\DeclareRobustCommand{\legsimcirc}{\mkcirc{legendgray}}
\DeclareRobustCommand{\legsimdia}{\mkdia{legendgray}}
\DeclareRobustCommand{\legsimtri}{\mktri{legendgray}}

\DeclareRobustCommand{\legthdash}{\tikz[baseline=-0.6ex]{%
  \draw[legendgray, line width=1.2pt, dash pattern=on 3pt off 2pt]
  (0,0) -- (13pt,0);}}
\DeclareRobustCommand{\legthdot}{\tikz[baseline=-0.6ex]{%
  \draw[legendgray, line width=1.2pt, dash pattern=on 1.2pt off 1.5pt]
  (0,0) -- (12pt,0);}}
\begin{document}
%\setstcolor{red}
\title{ Anomalous Mobility Enhancement in Restricted Geometries
}

\author{Yuval Zippin\orcidlink{0009-0009-2422-7430}}
\email{zippinyuval@gmail.com}
\author{Stanislav Burov\orcidlink{0000-0003-1065-174X}}

\email{stasbur@gmail.com}
\affiliation{Department of Physics, Bar-Ilan University, Ramat-Gan 5290002,
Israel}

%\author{R. K. Singh}
%\email{rksinghmp@gmail.com}
%\affiliation{Department of Physics, Bar-Ilan University, Ramat-Gan 5290002, Israel}
%\affiliation{Department of Biomedical Engineering, Ben-Gurion University of the Negev, Be’er Sheva 85105, Israel}

%\author{Stanislav Burov}
%\email{stasbur@gmail.com}
%\affiliation{Department of Physics, Bar-Ilan University, Ramat-Gan  5290002, Israel}

\begin{abstract}
We investigate the effect of geometrical confinement on
driven transport in anomalous media within the framework
of the quenched trap model.
The particle performs a biased random walk on a lattice,
where quenched disorder introduces power-law distributed trapping times.
We consider channels that are infinite along one direction
and have a finite cross-section in the remaining $d-1$ directions.
In the glassy regime, where $\alpha=T/T_g<1$,
we derive an analytical expression for the mean longitudinal
displacement in the long-time and weak-force limits.
We find the scaling
$\langle x_\parallel(t)\rangle
\propto F_\parallel^\alpha t^\alpha
\Omega^{-(1-\alpha)}$,
where $F_\parallel>0$ is the effective longitudinal force
and $\Omega$ is the number of sites
in a transverse cross-section. 
Thus, the anomalous mobility is enhanced for narrower channels.
Our analytical result applies to general Bravais lattices in $d$ dimensions 
with periodic transverse boundary conditions.
Numerical simulations confirm the analytical predictions for different lattice geometries and show quantitative agreement even for channels with reflecting boundary conditions.

\end{abstract}

\maketitle

\newcommand{\fr}{\frac}
\newcommand{\lr}{\langle}
\newcommand{\rl}{\rangle}
\newcommand{\tl}{\tilde}

\section{Introduction}
\label{sec:Introduction}
Situations in which a phenomenon contradicts our intuition usually lead us to recognize that our understanding of nature was once again too naive.
Enhanced transport properties due to increased confinement is one such example.
While it is traditionally assumed that increasing confinement, such as narrowing a channel cross section, will also slow motion in the medium, this need not be the case. 
Especially when dealing with disordered or amorphous material or nonspherical particles. 
The motion of molecules inside zeolites shows the so-called transport anomaly: when the zeolite's cross-section is narrowed down, the diffusivity can increase~\cite{bandyopadhyay1995diffusion,yuan2023hyperloop,brandani2024diffusion,qi2026confinement}.
A similar diffusivity-enhancement effect was observed for rod-like particles in a macromolecular network~\cite{yang2025nonmonotonic} and for worm and active-polymer translocation in quasi-1D channels~\cite{prathyusha2026active}. 
Recently, one of us reported mobility enhancement due to channel narrowing for a driven particle in a model of amorphous media~\cite{Shafir2024}. 
While the details of the systems where such transport enhancement was observed differ, the general idea behind the phenomenon is similar: by decreasing confinement, we allow the system to explore additional configurations. 
Some of these configurations negatively affect transport. 
When the effect is drastic, these additional degrees of freedom slow the motion compared to the more confined setup.

In this work, we focus on the enhancement of transport properties, as it appears for driven particles in amorphous materials. 
Specifically, we consider the quenched trap model~\cite{Bouchaud1990} that was originally constructed to mimic glassy behavior.  
Dynamical arrest, which is usually present in glassy systems, was introduced into QTM as local energetic traps with diverging mean escape time. 
The motion in QTM, as specified in the next section, is in a disordered landscape where quenched local traps strongly suppress transport. 
This combination of quenched disorder and diverging dwell times is what made analytical exploration of this model very challenging. 
Numerical simulations and scaling arguments revealed early on the presence of anomalous diffusion in QTM when the temperature was sufficiently low~\cite{bernasconi1980diffusion,RevModPhys.53.175,Machta1985}.
The process was explored under the semi-equilibrium assumptions~\cite{derrida1983velocity} and various simplifications such as strictly directed dynamics~\cite{aslangul1990dynamical}, 
and the exact representation of the positional probability density function (PDF) was first obtained in the limit of very low temperatures~\cite{PhysRevE.68.036114}. 
One of the interesting effects that are present in QTM but not in other models of anomalous diffusion like the continuous time random walk~\cite{Bouchaud1990} is the non-linear response to external drive~\cite{PhysRevE.69.026103,Burov2020,Shafir2022}. 
It was found that when the motion occurs in 1D and an external driving force is present, the position of the particle grows sublinearly with the external force $f$, even in the weak force limit. 
Although the QTM is almost $40$ years old, it remains one of the central models of anomalous diffusion~\cite{araujo2026sub,shirataki2026observation}. 
The mentioned observation of mobility enhancement in channels~\cite{Shafir2024} when the dynamics follows the QTM is just one additional example of why this model still attracts scientific interest. 

The work below generalizes the results from \cite{Shafir2024}, which were obtained for a specific 2D channel with a square lattice. 
Here we study the response to an external drive in the anomalous phase, while the particle is confined to a channel with arbitrary geometry and dimensionality. 
The new results hold for any type of Bravais lattice. 
The main effect remains similar to what was observed in~\cite{Shafir2024}: the mobility increases as the channel cross section decreases. 
Our explicit calculation of the return probability of a random walk in the presence of a weak external force, for a general lattice and confinement, is what allows us to obtain the analytic form for the mean particle position in QTM. The results are expressed in terms of the diffusion tensor of the underlying unbiased spatial random walk.
The combination of the quenched disorder and scale-free property of the trapping times is crucial. Without either ingredient, the effect disappears. 

The manuscript is organized as follows.
In Sec.~\ref{sec:Model}, we introduce the QTM and derive the general expression for the mean
longitudinal displacement using the double-subordination
approach.
In Sec.~\ref{sec:ReturnProb}, we calculate the leading
weak-force behavior of the return probability for confined
random walks on general Bravais lattices with periodic
boundary conditions.
In Sec.~\ref{sec:Results}, we derive the dependence of
anomalous mobility on the applied force and confinement,
and compare the analytical predictions with numerical
simulations for different lattice geometries and boundary
conditions.
In Sec.~\ref{sec:Discussion} we discuss the results and observed phenomena. 
Finally, Appendix~\ref{app:zero_mode} provides the details
of the zero-mode integral evaluated in
Sec.~\ref{sec:ReturnProb}.

\section{The Quenched Trap Model}
\label{sec:Model}
The QTM describes a thermally activated particle hopping between discrete
energetic traps~\cite{Bouchaud1990}. A trap located at position $\mathbf{r}$ is
characterized by its depth $E_{\mathbf{r}}>0$, and escape from it is an activated
process, so that the time the particle spends there is given by the Arrhenius law
$\tau_{\mathbf{r}} \propto \exp(E_{\mathbf{r}}/T)$, where $T$ is the temperature
and the Boltzmann constant is set to $k_B=1$~\cite{Monthus1996}. The disorder of
the medium is carried entirely by the depths, which are taken to be exponentially
distributed, $\phi(E)=\exp(-E/T_g)/T_g$, with $T_g$ the glass transition
temperature. Since the Arrhenius law is exponential in the depth, it maps this
exponential distribution onto a power-law distribution of the dwell times, the tail of which
reads~\cite{Burov2020, Shafir2024}:
\begin{equation}
    \psi(\tau_{\mathbf{r}}) \sim
    \frac{A}{|\Gamma(-T/T_g)|}\,\tau_{\mathbf{r}}^{-1-(T/T_g)},
    \qquad (\tau_{\mathbf{r}} \to \infty),
    \label{eq:waiting_time_pdf}
\end{equation}
Here $\Gamma(\dots)$ is the Gamma function and $A > 0$ is the amplitude of the
distribution. The tail is written in this particular form so that the small-$u$
expansion of the Laplace transform of Eq.~(\ref{eq:waiting_time_pdf}) reads
$\hat{\psi}(u) \sim 1 - A u^{T/T_g}$ for small $u$.
%and it is through this coefficient that the
%disorder will enter the derivation below.

Since the temperature enters only through the ratio $T/T_g$, we denote it by the
dimensionless parameter $\alpha = T/T_g$ and use it from here on. The mean escape
time follows from the tail of Eq.~(\ref{eq:waiting_time_pdf}) through
$\langle \tau \rangle = \int_0^{\infty} \tau \psi(\tau)\,d\tau$, and in the
low-temperature glassy regime where $0 < \alpha < 1$, this integral diverges,
$\langle \tau \rangle \to \infty$. 
The dwell times then have no characteristic
scale, and the model displays aging, weak ergodicity breaking and
non-self-averaging~\cite{Bouchaud1992, Monthus2003, Akimoto2018,
bamb2025concentration}. This is the regime considered throughout this work.

The dwell times are assigned to the lattice before the motion begins. Every site
$\mathbf{r}$ carries a single waiting time $\tau_{\mathbf{r}}$, drawn
independently from Eq.~(\ref{eq:waiting_time_pdf}), so that the set
$\{\tau_{\mathbf{r}}\}$ is a collection of positive, independent and identically
distributed random variables that constitutes a fixed realization of the
disorder~\cite{Arous2015, Cerny2015}. This is what is meant by quenched disorder,
and it is the sole difference between the QTM and the continuous time random walk
(CTRW), where a new dwell time is generated at every revisit of a lattice site. 
In QTM, the particle that
returns to a site it has already visited is localized there for exactly the same
duration $\tau_{\mathbf{r}}$ as before~\cite{Burov2011, Burov2012, Burov2017}, and
the total time that is accumulated during the motion of the particle is tied to the trajectory it traces rather than
to the number of steps it takes. 
The dwell times remain independent from site to
site, but the times the particle actually experiences do not, and this correlation
significantly raises the difficulty bar of analytical treatment of QTM. 

We assume that the spatial motion takes place on a $d$-dimensional Bravais
lattice~\cite{Machta1985, Klafter2011}, whose basis vectors
$\{\mathbf{a}_\nu\}$ need not be of equal length, each direction carrying its own
lattice constant $a_\nu = |\mathbf{a}_\nu|$. The particle starts at the origin
$\mathbf{r} = \mathbf{0}$ at $t = 0$, waits there for the time
$\tau_{\mathbf{0}}$ assigned to that site, and then hops to one of its nearest
neighbours. It is trapped at the new site for the time
stored there, and the process repeats. The probability of a hop from
$\mathbf{r}$ to $\mathbf{r}'$ is $p(\mathbf{r}'; \mathbf{r})$, and since the
lattice is translationally invariant it depends on the displacement alone,
$p(\mathbf{r}'; \mathbf{r}) = p(\mathbf{r}' - \mathbf{r})$~\cite{Montroll1965,
Hughes1995}.

The quantity we wish to obtain is the disorder-averaged propagator
$\langle P(\mathbf{r},t) \rangle$, the probability of finding the particle at
$\mathbf{r}$ after a measurement time $t$, averaged over all realizations of the
trapping landscape.
%Writing it down directly is obstructed by the very correlation described above. 
The position of the particle is determined by the sequence of
jumps it performs, while the time it takes to perform them is determined by the
sites it happens to land on and by how often it returns to each of them, so that
the spatial and the temporal parts of the process cannot be treated as
independent~\cite{Burov2017}. 
%The route around this difficulty is not to separate space from time directly, but to interpose between them a clock of the process itself, a local time that counts how far the walk has advanced rather than how long it has taken. 
The route around this difficulty is to establish a local time for the process that is independent of temporal disorder and defines the spatial motion.
The propagator is then assembled in two stages, a construction
known as double subordination~\cite{Burov2020, Shafir2022}. 
In the first stage, the
measurement time $t$ is related to this local time, which is where the disorder
enters and where the correlations are absorbed. 
In the second stage, the local time is expressed in terms of the number of jumps $N$.
This second stage maps the process to an ordinary random walk on the lattice, free of disorder. 
Below, we redevelop the construction of these two stages following the general prescription for transient QTM established in ~\cite{Burov2020} and applied
to biased restricted geometries in Ref.~\cite{Shafir2024}.

The first stage asks how much time has elapsed once the walker has traced a given path, and the answer is fixed by the landscape it has crossed. Each arrival at a site costs the dwell time stored there, so that the measurement time accumulated along the trajectory is:
\begin{equation}
    t = \sum_{\mathbf{r}} n_{\mathbf{r}} \tau_{\mathbf{r}},
    \label{eq:measurement_time}
\end{equation}
where $n_{\mathbf{r}}$ counts the arrivals at $\mathbf{r}$ up to $t$~\cite{Burov2007}. The two factors in this sum are of a different nature. The dwell times $\tau_{\mathbf{r}}$ are independent from site to site, while the visitation numbers $n_{\mathbf{r}}$ are not, since the walk moves between nearest neighbours and cannot visit a site many times without visiting its neighbours a comparable number of times. We therefore consider the $\{n_{\mathbf{r}}\}$ as
fixed, the outcome of one given trajectory, and average over the disorder alone, that is over the realization $\{\tau_{\mathbf{r}}\}$. The average is examined in
the Laplace domain, where the sum in the exponent becomes a product over the visited sites, and where the independence of the dwell times lets each factor be averaged separately, so that $\langle e^{-ut} \rangle = \prod_{\mathbf{r}}
\hat{\psi}(n_{\mathbf{r}}u)$ with $\hat{\psi}(u) = \int_0^\infty
e^{-u\tau}\psi(\tau)\,d\tau$. The long-time behaviour corresponds to $u \to 0$,
where Eq.~(\ref{eq:waiting_time_pdf}) supplies $\hat{\psi}(u) = 1 - Au^\alpha$,
and each factor of the product becomes $1 - A n_{\mathbf{r}}^\alpha u^\alpha$.
Taking the logarithm turns the product back into a sum, and the leading term of the expansion is $-Au^\alpha \sum_{\mathbf{r}} n_{\mathbf{r}}^\alpha$, all the remaining terms being shown in Ref.~\cite{Burov2020} to vanish in the limit considered here. In the long-time limit we therefore have~\cite{Burov2011, Burov2012}:
\begin{equation}
    \langle e^{-ut} \rangle \simeq e^{-A S_\alpha u^\alpha},
    \qquad
    S_\alpha = \sum_{\mathbf{r}} (n_{\mathbf{r}})^\alpha .
    \label{eq:local_time}
\end{equation}
The trajectory has thus entered the temporal statistics through a single number. Where the walker went and in what order doesn't matter; only the combination $S_\alpha$ survives the average. 
It is $S_\alpha$ that plays the role of the local time of the process, the clock to which the motion is subordinated instead of $t$. 
The right-hand side of Eq.~(\ref{eq:local_time}) is the Laplace transform of the one-sided L\'evy stable distribution $l_{\alpha,A,1}$~\cite{Metzler2000}, and this is what allows us to establish the first, i.e., the temporal, stage. 
For a given local time $S_\alpha$, the
measurement time is the random variable $t(S_\alpha) = S_\alpha^{1/\alpha}\eta$,
with $\eta$ distributed according to $l_{\alpha,A,1}(\eta)$. Any experiment,
however, prescribes the opposite. It is $t$ that is set in advance and $S_\alpha$
that is left to fluctuate, so the roles of parameter and random variable must be
exchanged. The exchange is permitted because $S_\alpha$ grows strictly with every
jump the particle performs, so that its relation to $t$ is one to one and can be
inverted, and it yields the probability density of observing a given local time at a fixed measurement time~\cite{Burov2011, Burov2012}:
\begin{equation}
    \mathcal{Q}_t(S_\alpha) \sim \frac{t}{\alpha}\, S_\alpha^{-1/\alpha - 1}\,
    l_{\alpha,A,1} \left( \frac{t}{S_\alpha^{1/\alpha}} \right).
    \label{eq:S_alpha_pdf}
\end{equation}
Equation~(\ref{eq:S_alpha_pdf}) is the disorder-averaged link between the two
clocks, and it completes the first subordination stage. The propagator now follows from conditional probability alone. Let $P_{S_\alpha}(\mathbf{r})$ be the probability of finding the particle at $\mathbf{r}$ once a local time $S_\alpha$ has elapsed, a purely spatial quantity that carries no disorder, since it depends on the visitation numbers and not on the times stored at the sites. Whenever the particle is found at $\mathbf{r}$ at
time $t$, some value of $S_\alpha$ must have been reached, so summing over all of
them with the weight of Eq.~(\ref{eq:S_alpha_pdf}) gives the disorder-averaged
propagator~\cite{Burov2011, Burov2012}:
\begin{equation}
    \langle P(\mathbf{r},t) \rangle
    = \sum_{S_\alpha} P_{S_\alpha}(\mathbf{r})\, \mathcal{Q}_t(S_\alpha).
    \label{eq:first_subordination}
\end{equation}
What Eq.~(\ref{eq:first_subordination}) leaves open is $P_{S_\alpha}(\mathbf{r})$
itself, which is still not an ordinary random walk, since the local time advances
by an amount that depends on how often the sites were visited during the motion. 
The next step is to subordinate once more, now to the number of jumps $N$.
%for which the walk is elementary. 
Let $W_N(\mathbf{r})$ be the probability of
finding the particle at $\mathbf{r}$ after exactly $N$ jumps, and
$\mathcal{G}_{S_\alpha}(N,\mathbf{r})$ the probability that
the prescribed local time $S_\alpha$ is reached at the $N$-th jump, conditioned
on the particle being at $\mathbf{r}$ after those $N$ jumps.
Summing over all possible $N$s gives $P_{S_\alpha}(\mathbf{r}) = \sum_{N=0}^{\infty}
W_N(\mathbf{r}) \mathcal{G}_{S_\alpha}(N,\mathbf{r})$, and inserting this into
Eq.~(\ref{eq:first_subordination}) completes the double-subordination
representation of the propagator~\cite{Burov2020, Shafir2022}:
\begin{equation}
    \langle P(\mathbf{r},t) \rangle = \sum_{S_\alpha} \sum_{N=0}^{\infty}
    W_N(\mathbf{r})\, \mathcal{G}_{S_\alpha}(N, \mathbf{r})\,
    \mathcal{Q}_t(S_\alpha).
    \label{eq:double_subordination}
\end{equation}
The relation $t = S_\alpha^{1/\alpha}\eta$ fixes the mapping between the local time and the 
measure time. 
The random variable $\eta$ does not depend on $t$, so the
weight of $\mathcal{Q}_t(S_\alpha)$ sits on local times of order
$S_\alpha \sim t^\alpha$. 
%This scale grows without bound with the measurement time, leaving any fixed value of $S_\alpha$ behind, and 
The limit
$t \to \infty$ may therefore be taken as the limit $S_\alpha \to \infty$.  It remains to find
$\mathcal{G}_{S_\alpha}(N,\mathbf{r})$ in that limit. 
For a fixed number of jumps $N$, the local time varies from one trajectory to another.
But when the
walk is transient, that is when the probability of eventual return to the
origin $Q_0$ is strictly smaller than one, $Q_0 < 1$~\cite{Weiss1994, Hughes1995},
these fluctuations become negligible. 
In the limit when  $N\to\infty$, the typical $S_\alpha$ not only diverges but
concentrates on a single value proportional to $N$~\cite{Burov2020}. 
So in the limit $N \to \infty$ the two variables, $S_\alpha$ and $N$,  are
interchangeable, independently of the position $\mathbf{r}$,
\begin{equation}
    \mathcal{G}_{S_\alpha}(N,\mathbf{r}) \longrightarrow
    \delta\!\left(S_\alpha/\Lambda -  N\right),
\end{equation}
with the constant $\Lambda$ fixed by the eventual return probability $Q_0$ 
~\cite{Burov2017}:
\begin{equation}
    \Lambda = \frac{(1-Q_0)^2}{Q_0}\,\mathrm{Li}_{-\alpha}(Q_0).
    \label{eq:lambda}
\end{equation}
where $\mathrm{Li}_a(b) = \sum_{j=1}^\infty b^j / j^a$ is the polylogarithm
function~\cite{Abramowitz1988, Gradshteyn2014}. 
%The geometry of the channel and the external drive enter the ropagator through $\Lambda$ alone, and they do so only by way of the return probability $Q_0$, which is fixed by the spatial jump probabilities and is evaluated for our channel in Sec.~\ref{sec:ReturnProb}. 
Replacing $\mathcal{G}_{S_\alpha}(N,\mathbf{r})$
by the delta function carries out the sum over $S_\alpha$ in
Eq.~(\ref{eq:double_subordination}), and setting $S_\alpha = \Lambda N$ in
$\mathcal{Q}_t(S_\alpha)$ leaves the asymptotic form of the disorder-averaged
propagator~\cite{Burov2011, Burov2017, Burov2020}:
\begin{equation}
    \langle P(\mathbf{r},t) \rangle \sim \sum_N W_N(\mathbf{r})
    \frac{t/\Lambda^{1/\alpha}}{\alpha N^{1/\alpha + 1}}
    l_{\alpha,A,1}\!\left( \frac{t/\Lambda^{1/\alpha}}{N^{1/\alpha}} \right).
    \label{eq:final_propagator}
\end{equation}
Equation~(\ref{eq:final_propagator}) expresses the macroscopic propagator of
the QTM through the discrete spatial random walk $W_N(\mathbf{r})$ in the
asymptotic long-time limit. 
The constant $\Lambda$ carries the effect of the
lattice geometry and of the external force. 
Eq.~\eqref{eq:lambda} shows that $\Lambda$ is defined only by the eventual return
probability $Q_0$. 
%Therefore, all the effect of external driving and geometry is encapsulated in $Q_0$.
Setting $\Lambda = 1$ reduces the expression to the propagator of the continuous time random walk~\cite{Bouchaud1990, Metzler2000}. 
Thus we can say  that $\Lambda$ measures the departure of the quenched dynamics from the case of annealed disorder where a new dwell time is drawn at every
visit~\cite{Burov2011, Burov2012, Burov2017}. 
Since $Q_0$ is fixed by the spatial random walk, it is the geometry of the
lattice on which that walk takes place that remains to be specified. 
We consider a channel of finite cross-section~\cite{Malgaretti2019,
Benichou2018}, built on a $d$-dimensional Bravais lattice. The motion is
unrestricted along a single longitudinal axis $\hat{{\bf n}}$, while the remaining $d-1$ directions are
bounded, each transverse axis $j$ spanning $w_j$. 
The assumption of translational invariance requires the transition
probabilities to depend on the displacement alone, $p(\mathbf{r}' -
\mathbf{r})$, and we therefore close the bounded directions by periodic
boundary conditions~\cite{Hughes1995}. 
The case of reflecting walls is
addressed in Sec.~\ref{sec:Results}.

The confinement enters the transport through the number of sites contained in a transverse periodic cell, denoted by $\Omega$ (see Sec.~\ref{sec:ReturnProb}).
In the simple
hexagonal channel considered in  Sec.~\ref{sec:Results} the cross-section is a hexagonal
patch of a triangular layer, containing $w$ lattice sites along each side and a total of $\Omega = 3w^2-3w+1$
sites. 
As shown in Sec.~\ref{sec:ReturnProb}, the leading weak-force dependence of $Q_0$ on confinement is determined by $\Omega$, rather than by the individual transverse widths. 

With the geometry fixed, the only remaining choice is the drive. An
external force $\mathbf{f}$ is applied to the particle, and it acts on the
transition probabilities rather than on the dwell times, which stay as assigned by the disorder ~\cite{Bertin2003response, Akimoto2020}. 
The strength of $\mathbf{f}$ is measured against the temperature through the effective force $\mathbf{F} = \mathbf{f}/T$, the Boltzmann constant being set to $k_B=1$~\cite{Bouchaud1990}. Jumps are restricted to nearest neighbours along the vectors $\mathbf{e}_\nu$, and detailed balance fixes the ratio of the probabilities of opposite jumps to $p(\mathbf{e}_\nu)/p(-\mathbf{e}_\nu) = e^{\mathbf{F} \cdot
\mathbf{e}_\nu}$~\cite{Akimoto2020, Shafir2022}, so that the drive favours the direction it points along and suppresses the reverse one. Together with the normalization $\sum_\nu p(\mathbf{e}_\nu) = 1$ this determines the jump probabilities completely:
\begin{equation}
    p(\mathbf{e}_\nu) = \mathcal{N} e^{\mathbf{F} \cdot \mathbf{e}_\nu / 2},
    \label{eq:jump_prob}
\end{equation}
with $\mathcal{N} = \left[ \sum_{\nu'} e^{\mathbf{F} \cdot \mathbf{e}_{\nu'}
/ 2} \right]^{-1}$. Without the applied force, all the allowed jumps $\mathbf{e}_\nu$ are equally probable. 
%We take the force along the longitudinal axis, leaving the transverse jump probabilities symmetric. 

The observable we wish to study is the mean displacement along the unrestricted
longitudinal axis, $\langle x_\parallel(t) \rangle = \sum_{\mathbf{r}}
(\mathbf{r} \cdot \hat{{\bf n}}) \langle P(\mathbf{r},t)
\rangle$~\cite{Burov2020, Shafir2022}. The jump probabilities of
Eq.~(\ref{eq:jump_prob}) supply it at once. A single jump advances the walker
along the channel by $v_\parallel = \hat{{\bf n}} \cdot \sum_{\nu}
\mathbf{e}_\nu p(\mathbf{e}_\nu)$ on average, the longitudinal drift velocity,
and since successive jumps are independent, the mean position after $N$ steps
is $\sum_{\mathbf{r}} (\mathbf{r} \cdot\hat{{\bf n}}) W_N(\mathbf{r}) = N
v_\parallel$. 
The spatial part of the propagator therefore responds linearly
to $v_\parallel$ and all that is left to perform is the transformation from $N$ to the measurement time.
Inserting $N v_\parallel$ into Eq.~(\ref{eq:final_propagator}) gives:
\begin{equation}
    \langle x_\parallel(t) \rangle \sim \sum_N N v_\parallel
    \frac{t/\Lambda^{1/\alpha}}{\alpha N^{1/\alpha + 1}} l_{\alpha,A,1}
    \left( \frac{t/\Lambda^{1/\alpha}}{N^{1/\alpha}} \right).
    \label{eq:mean_disp_sum}
\end{equation}
In the long-time limit ($t \to \infty$) the sum over $N$ in
Eq.~(\ref{eq:mean_disp_sum}) is replaced by an integral~\cite{Burov2011}, and
the dependence on $N$ is removed by the change of variables $y =
t/(\Lambda N)^{1/\alpha}$, the same ratio that appears in the argument of the
L\'evy density. 
The Jacobian $dN = \alpha (t/\Lambda^{1/\alpha})^{\alpha}
y^{-\alpha-1} dy$ cancels the prefactor of Eq.~(\ref{eq:mean_disp_sum}) up to
a single power of $y$, and what is left is a fractional moment of the L\'evy
distribution:
\begin{equation}
    \langle x_\parallel(t) \rangle \sim v_\parallel
    \left( \frac{t}{\Lambda^{1/\alpha}} \right)^{\alpha}
    \int_0^\infty y^{-\alpha} \, l_{\alpha,A,1}(y) \, dy .
    \label{eq:levy_moment}
\end{equation}
The moment is the one of order $q = -\alpha$, and it is
finite~\cite{Burov2011, Burov2012}. Evaluating it through the identity
$\int_0^\infty y^{q} l_{\alpha,A,1}(y) dy = A^{q/\alpha}\Gamma(1 -
q/\alpha)/\Gamma(1 - q)$, valid for $q/\alpha< 
1$~\cite{Barkai2001}, gives $1/[A\Gamma(1+\alpha)]$, while the prefactor of
Eq.~(\ref{eq:levy_moment}) is $t^{\alpha}/\Lambda$. Substituting $\Lambda$
from Eq.~(\ref{eq:lambda}) leaves the asymptotic longitudinal
displacement~\cite{Burov2020, Shafir2022}:
\begin{equation}
    \langle x_\parallel(t) \rangle \sim \frac{v_\parallel}{A \Gamma(1+\alpha)}
    \frac{Q_0}{(1-Q_0)^2 \mathrm{Li}_{-\alpha}(Q_0)} \, t^\alpha .
    \label{eq:mean_position}
\end{equation}
The displacement grows as $t^\alpha$ rather than linearly in time, reflecting the broad distribution of trapping times in the glassy regime~\cite{Barkai1998, Shafir2024}. The drive enters through $v_\parallel$, and the lattice and the confinement enter through the return
probability $Q_0$ alone~\cite{Burov2017, Burov2020}. 

Equation \eqref{eq:mean_position} shows that $ Q_0$ encapsulates the nontrivial dependence on the force and geometry. 
To capture the effects of confinement and geometry, we must obtain the explicit form of $Q_0$.
We now turn to
evaluating this quantity.

\section{The Return Probability}
\label{sec:ReturnProb}

To complete the physical picture established in the previous section, our goal is to evaluate the eventual return probability $Q_0$ and derive its leading dependence on the transverse cross-section for a weak external force. 
To find $Q_0$, we must relate the probability of finding the particle at the origin to its first-return probability. We solve this relationship by introducing generating functions and transitioning to $k$-space to account for the confined lattice geometry~\cite{Montroll1965, Weiss1994, Hughes1995}. 

Consider a random walk starting at the origin, with initial occupation probability $W_0(\mathbf{r}) = \delta_{\mathbf{r}, \mathbf{0}}$. The probability $W_N(\mathbf{0})$ of finding the walker at the origin after $N$ steps is related to the first-return probability, $f_N(\mathbf{0})$, which is the probability of returning to the origin for the exact first time at step $N \ge 1$. To establish the connection between these quantities, we follow the trajectory of the walker. For the walk to be at the origin at step $N \ge 1$, it must first reach the origin at some earlier step $i$ ($1 \le i \le N$). Then, the walk must return to the origin in the remaining $N-i$ steps. Since this latter process has a probability $W_{N-i}(\mathbf{0})$, the probability for this sequence is the product $f_i(\mathbf{0}) W_{N-i}(\mathbf{0})$. Summing over all possible first-return times $i$, and incorporating the initial condition via a Kronecker delta, gives the discrete renewal equation~\cite{Feller1968, Redner2001, Benichou2011}: 
\begin{equation}
    W_N(\mathbf{0}) = \delta_{N,0} + \sum_{i=1}^N f_i(\mathbf{0}) W_{N-i}(\mathbf{0}).
\end{equation}
Convolution equations of this form are most naturally solved by introducing generating functions, which serve as the discrete analog of the Laplace transform: $\tilde{W}_z(\mathbf{0}) = \sum_{N=0}^\infty W_N(\mathbf{0}) z^N$ and $\tilde{f}_z(\mathbf{0}) = \sum_{N=1}^\infty f_N(\mathbf{0}) z^N$. By multiplying the renewal equation by $z^N$ and summing over all $N$, the convolution theorem for power series transforms the discrete sum into a direct product of the generating functions~\cite{Weiss1994, Hughes1995}, yielding $\tilde{W}_z(\mathbf{0}) = 1 + \tilde{f}_z(\mathbf{0}) \tilde{W}_z(\mathbf{0})$. Rearranging this expression provides the first-return generating function, $\tilde{f}_z(\mathbf{0}) = 1 - 1/\tilde{W}_z(\mathbf{0})$. The eventual return probability to the origin, $Q_0$, is the sum of the first-return probabilities over all steps, $Q_0 = \sum_{N=1}^\infty f_N(\mathbf{0})$. This sum is recovered by evaluating the generating function in the limit $z \to 1^-$, yielding P\'olya's classical result~\cite{ Hughes1995}:

\begin{equation}
    \label{eq:polya}
    Q_0 = 1 - \lim_{z \to 1^-} \frac{1}{\tilde{W}_z(\mathbf{0})}.
\end{equation}
Equation~\eqref{eq:polya} provides a general framework to determine $Q_0$, but its value depends on the specific lattice geometry and the external force. To incorporate these physical constraints, we must evaluate $\tilde{W}_z(\mathbf{0})$ by introducing the characteristic function of the single-step distribution:

\begin{equation}
    \lambda(\mathbf{k}) = \sum_{\Delta\mathbf{r}} p(\Delta\mathbf{r}) e^{i \mathbf{k} \cdot \Delta\mathbf{r}}.
    \label{eq:lambdaDef}
\end{equation}
where $\Delta\mathbf{r}$ represents the set of discrete lattice vectors connecting a given site to its accessible neighbors, and $p(\Delta\mathbf{r})$ denotes the corresponding transition probabilities. By the discrete convolution theorem, the Fourier transform of the $N$-step probability distribution factorizes as $\widehat{W}_N(\mathbf{k}) = \lambda(\mathbf{k})^N$~\cite{Montroll1965, Weiss1994}.

Since the random walk takes place on a Bravais lattice, we
express the position of each site as
$\mathbf{r}=\sum_{j=1}^{d}n_j\mathbf{a}_j$, where
$\{\mathbf{a}_j\}$ are the primitive lattice vectors and
$n_j$ are integers. These vectors need not be orthogonal
or have equal lengths. All sites of a Bravais lattice are
equivalent under lattice translations. Since the transition
probabilities depend only on the displacement between sites,
the random walk is translationally invariant, as required
for the mapping introduced in Sec.~\ref{sec:Model}.
The wave vectors $\mathbf{k}$ are restricted to the first
Brillouin zone (BZ) of the lattice. Substituting the inverse
Fourier transform into the definition of the generating
function and summing the resulting geometric series over
$N$, we obtain the generating function at the origin~\cite{Hughes1995}:
%%%%%%%%%%%%%%%%%
\begin{equation}
\widetilde W_z(0)=
\frac{V_{\mathrm{cell}}}{(2\pi)^d}
\int_{\mathrm{BZ}}
\frac{d^d k}{1-z\lambda(\mathbf{k})},
\label{eq:Wz0}
\end{equation}
%%%%%%%%%%%%%%%%%%
where $V_{\mathrm{cell}}=
|\det(\mathbf{a}_1,\ldots,\mathbf{a}_d)|$
is the volume of a primitive lattice cell.

As introduced in Sec.~\ref{sec:Model}, we consider a straight
channel that extends infinitely along its longitudinal axis
$\hat{\mathbf n}$ and has a finite cross-section in the remaining
$d-1$ directions. For a general Bravais lattice, we choose the
channel orientation such that its longitudinal axis is perpendicular
to $d-1$ primitive lattice vectors, which we take to be
$\mathbf a_2,\ldots,\mathbf a_d$. Thus,
$\hat{\mathbf n}\cdot\mathbf a_j=0$ for $j=2,\ldots,d$.
Periodic boundary conditions are imposed by identifying lattice
sites separated by $w_j\mathbf a_j$, where $w_j$ is a positive
integer. A particle crossing a transverse boundary therefore
re-enters the channel at an equivalent lattice site on the
opposite side. Since these identifications correspond to lattice
translations, the local connectivity and transition probabilities
remain unchanged, preserving translational invariance.
The number of distinct lattice sites in each longitudinal repeat is
%%%%%%%%%%%%%%%%%%%%%%%%%%%%
\begin{equation}
\Omega=\prod_{j=2}^{d}w_j.
\label{eq:OmegaDefinition}
\end{equation}
%%%%%%%%%%%%%%%%%%%%%%%%%%%

The periodic identification of lattice sites allows us to
construct the probability distribution in the confined channel
from that of the unrestricted lattice. A site at position
$\mathbf r$ inside the channel is equivalent to all sites
displaced from it by integer multiples of the transverse
periodicity vectors $w_j\mathbf a_j$. Since these translations
are perpendicular to the longitudinal axis, they leave the
longitudinal coordinate unchanged,
$\hat{\mathbf n}\cdot
\left(\sum_{j=2}^{d}m_jw_j\mathbf a_j\right)=0$.
Consequently, the probability $W_N^*(\mathbf r)$ of finding
the particle at $\mathbf r$ after $N$ steps in the confined
channel is obtained by summing the unrestricted probability
over all these equivalent positions:
%%%%%%%%%%%%%%%%%%%%%%%%%%%%%%%
\begin{equation}
W_N^*(\mathbf r)=
\sum_{m_2,\ldots,m_d\in\mathbb Z}
W_N\left(
\mathbf r+\sum_{j=2}^{d}m_jw_j\mathbf a_j
\right).
\label{eq:folded_probability}
\end{equation}
%%%%%%%%%%%%%%%%%%%%%%%%%%%%%%%%%
Transitioning to $k$-space, we take the Fourier transform
of the folded probability distribution in
Eq.~\eqref{eq:folded_probability}. Using the shift theorem,
each spatial translation $m_j w_j \mathbf a_j$ introduces
a phase factor $e^{-i m_j w_j \mathbf k\cdot\mathbf a_j}$.
The multiple sum over transverse translations can therefore
be written as
$\sum_{m_2,\ldots,m_d\in\mathbb Z}
e^{-i\sum_{j=2}^{d}m_j w_j\mathbf k\cdot\mathbf a_j}=\sum_{m_2,\ldots,m_d\in\mathbb Z}
\prod_{j=2}^{d}
e^{-i m_j w_j\mathbf k\cdot\mathbf a_j}$.
The summation indices are independent, allowing us to
exchange the multiple sum with the product.
Each factor is evaluated using the Poisson summation formula,
$\sum_{m=-\infty}^{\infty}e^{-imx}
=2\pi\sum_{\ell=-\infty}^{\infty}\delta(x-2\pi\ell)$.
Setting $x=w_j\mathbf k\cdot\mathbf a_j$,
the Fourier transform of the folded distribution becomes
%%%%%%%%%%%%%%%%%%%%%%%%%%%%%%%%%%%
\begin{equation}
\widehat W_N^*(\mathbf k)
=
\widehat W_N(\mathbf k)
\prod_{j=2}^{d}
\left[
\frac{2\pi}{w_j}
\sum_{\ell_j=-\infty}^{\infty}
\delta\left(
\mathbf k\cdot\mathbf a_j-
\frac{2\pi\ell_j}{w_j}
\right)
\right].
\label{eq:folded_fourier}
\end{equation}
%%%%%%%%%%%%%%%%%%%%%%%%%%%%%%%%%%%
To evaluate the generating function of the confined system,
we sum over $N$ and utilize the inverse Fourier representation of
Eq.~\eqref{eq:Wz0}.
Since the characteristic function is periodic under reciprocal
lattice translations,
$\lambda(\mathbf k+\mathbf G)=\lambda(\mathbf k)$,
the integration over the first Brillouin zone can be replaced
by an integration over any primitive cell of the reciprocal
lattice. We choose the parallelepiped generated by the
reciprocal primitive vectors $\mathbf b_j$, defined through
$\mathbf a_i\cdot\mathbf b_j=2\pi\delta_{ij}$.
Expressing the wave vector as
$\mathbf k=\sum_{j=1}^{d}t_j\mathbf b_j$,
the integration measure becomes
$d^dk=|\det(\mathbf b_1,\ldots,\mathbf b_d)|
\prod_{j=1}^{d}dt_j$.
Since
$|\det(\mathbf b_1,\ldots,\mathbf b_d)|
=(2\pi)^d/V_{\mathrm{cell}}$,
the Jacobian cancels the normalization prefactor in
Eq.~\eqref{eq:Wz0}.
The integration limits are
$-1/2\leq t_1<1/2$ and $0\leq t_j<1$
for $j=2,\ldots,d$. 
The transverse delta functions impose
$t_j=\ell_j/w_j$, so only the values
$\ell_j=0,\ldots,w_j-1$ contribute to the integral.
Integrating over the delta functions introduces a factor
$1/w_j$ for each transverse direction, yielding the
combined factor $1/\Omega$.
Since $\mathbf b_1$ is perpendicular to all transverse
primitive vectors, it is parallel to the channel axis
$\hat{\mathbf n}$.
Choosing the positive longitudinal direction such that
$a_\parallel=\hat{\mathbf n}\cdot\mathbf a_1>0$,
we obtain
$\mathbf b_1=(2\pi/a_\parallel)\hat{\mathbf n}$.
The remaining integration over $t_1$ can therefore be
expressed in terms of
$k_\parallel=2\pi t_1/a_\parallel$,
introducing an additional factor $a_\parallel/(2\pi)$.
The generating function of the confined system becomes
\begin{equation}
\begin{aligned}
&\widetilde W_z^*(0) =
\\
&\frac{a_\parallel}{2\pi\Omega}
\sum_{\ell_2=0}^{w_2-1}\cdots
\sum_{\ell_d=0}^{w_d-1}
\int_{-\pi/a_\parallel}^{\pi/a_\parallel}
\frac{dk_\parallel}
{1-z\lambda\left(
k_\parallel\hat{\mathbf n}
+\displaystyle\sum_{j=2}^{d}
\frac{\ell_j}{w_j}\mathbf b_j
\right)}.
\end{aligned}
\label{eq:Wz_confined}
\end{equation}
According to Eq.~\eqref{eq:polya}, the eventual return probability
$Q_0$ is determined by the limit $z\to1^-$ of the generating function.
We are interested in the weak-force regime, where the random walk
approaches recurrence and $Q_0\to1$.
Equation~\eqref{eq:polya} then implies that
$\widetilde W_1^*(0)$ diverges as the force vanishes.
To determine the leading behavior of this divergence, we examine
the denominator of Eq.~\eqref{eq:Wz_confined}.
Since the transition probabilities are positive and satisfy
the normalization condition
$\sum_{\Delta\mathbf r}p(\Delta\mathbf r)=1$,
Eq.~\eqref{eq:lambdaDef} implies that
$\lambda(\mathbf k)=1$ requires
$e^{i\mathbf k\cdot\Delta\mathbf r}=1$
for every allowed jump.
As the allowed jumps along the primitive lattice vectors
span the Bravais lattice, this condition is satisfied only
when $\mathbf k$ is a reciprocal lattice vector.
However, the integration and summations in
Eq.~\eqref{eq:Wz_confined} are restricted to a single
primitive reciprocal cell, where only $\mathbf k=0$
satisfies this condition.
Consequently, the divergence originates from the vicinity
of $\mathbf k=0$, while all other regions contribute finite
terms. Since $\mathbf k=0$ belongs to the transverse zero
mode, corresponding to $\ell_2=\cdots=\ell_d=0$,
we retain only this mode when evaluating the leading
weak-force behavior.
Namely, in the weak-force limit, we retain only the
transverse zero mode in Eq.~\eqref{eq:Wz_confined},
obtaining the leading contribution to the generating function:
%%%%%%%%%%%%%%%%%%%%%%%%%%%%%%%
\begin{equation}
\widetilde W_z^*(0)\simeq
\frac{a_\parallel}{2\pi\Omega}
\int_{-\pi/a_\parallel}^{\pi/a_\parallel}
\frac{dk_\parallel}
{1-z\lambda(k_\parallel\hat{\mathbf n})}.
\label{eq:Wz_zero_mode}
\end{equation}
%%%%%%%%%%%%%%%%%%%%%%%%%%%%%%%
To evaluate the leading weak-force behavior of Eq.~\eqref{eq:Wz_zero_mode}, we expand the characteristic function in
Eq.~\eqref{eq:lambdaDef} around $k_\parallel=0$ up to second order
%%%%%%%%%%%%%%%%%%%%%%%%%%%%%%
\begin{equation}
\lambda(k_\parallel\hat{\mathbf n})
\simeq 1+iv_\parallel k_\parallel-D_\parallel k_\parallel^2,
\label{eq:lambda_expansion}
\end{equation}
%%%%%%%%%%%%%%%%%%%%%%%%%%%%%%%
where the mean longitudinal displacement per step is
%%%%%%%%%%%%%%%%%%%%%%%%%%%%%%%
\begin{equation}
v_\parallel=
\sum_{\Delta\mathbf r}
p(\Delta\mathbf r)(\hat{\mathbf n}\cdot\Delta\mathbf r),
\label{eq:v_parallel}
\end{equation}
%%%%%%%%%%%%%%%%%%%%%%%%%%%%%%
and $D_\parallel=\frac12\sum_{\Delta\mathbf r}
p(\Delta\mathbf r)(\hat{\mathbf n}\cdot\Delta\mathbf r)^2$
is half the second moment of the longitudinal step
displacement. Substituting this expansion into the generating
function we obtain, for $z<1$,
%%%%%%%%%%%%%%%%%%%%%%%%%%%%%%
\begin{equation}
\widetilde W_z^*(0)\simeq
\frac{a_\parallel}{2\pi\Omega}
\int_{-\pi/a_\parallel}^{\pi/a_\parallel}
\frac{dk_\parallel}
{1-z+zD_\parallel k_\parallel^2
-izv_\parallel k_\parallel}.
\label{eq:Wz_quadratic}
\end{equation}
%%%%%%%%%%%%%%%%%%%%%%%%%%%%%%%%%
The integration in Eq.~\eqref{eq:Wz_quadratic} is evaluated
following the procedure detailed in Appendix~\ref{app:zero_mode}.
Taking the limit $z\to1^-$ at nonzero longitudinal drift,
we obtain the weak-force approximation
%%%%%%%%%%%%%%%%%%%%%%%%%%
\begin{equation}
\widetilde W_1^*(0)\simeq
\frac{a_\parallel}{\pi\Omega|v_\parallel|}
\left[
\frac{\pi}{2}
+\arctan\left(
\frac{\pi D_\parallel}
{a_\parallel|v_\parallel|}
\right)
\right].
\label{eq:W1_arctan}
\end{equation}
%%%%%%%%%%%%%%%%%%%%%%%%%%%
As $v_\parallel\to0$, the argument of the arctangent
diverges, and the leading contribution becomes
\begin{equation}
\widetilde W_1^*(0)\sim
\frac{a_\parallel}{\Omega|v_\parallel|}.
\label{eq:W1_leading}
\end{equation}
Although the arctangent expression also produces finite
terms upon expansion, these cannot be reliably determined
within our approximation. Indeed, finite contributions
arise both from the transverse modes neglected when
passing from Eq.~\eqref{eq:Wz_confined} to
Eq.~\eqref{eq:Wz_zero_mode}, and from regions of
reciprocal space where the small-${\bf k}$ expansion
of Eq.~\eqref{eq:lambda_expansion} is no longer justified.
We therefore retain only the leading divergent contribution.

Having obtained the leading weak-force behavior of the
generating function, we return to Eq.~\eqref{eq:polya},
which relates $\widetilde W_1^*(0)$ to the eventual return
probability $Q_0$. 
Substituting the asymptotic result
of Eq.~\eqref{eq:W1_leading}, we obtain that 
%%%%%%%%%%%%%%%%%%%%%%%%%%%%%
\begin{equation}
1-Q_0\simeq\frac{\Omega|v_\parallel|}{a_\parallel}.
\label{eq:Q0_weak}
\end{equation}
%%%%%%%%%%%%%%%%%%%%%%%%%%%%%%
This is the central result of this Section. 
It shows
that, in the weak-force regime, the probability of never
returning to the origin is proportional to the number
of lattice sites in the transverse periodic cell and to
the magnitude of the longitudinal drift.
The result holds for a fixed finite cross-section,
provided $\Omega|v_\parallel|/a_\parallel\ll1$,
and is independent of the detailed transverse geometry
at leading order.

To express the escape probability in terms of the applied
force, we use the Einstein relation for the underlying
lattice random walk. In the weak-force regime, with the
effective force $\mathbf F=F_\parallel\hat{\mathbf n}$
defined in Sec.~\ref{sec:Model}, the longitudinal drift
satisfies $v_\parallel\simeq D_0F_\parallel$~\cite{Kubo1966}, where $D_0$
is the longitudinal diffusion coefficient per step of
the unbiased walk. It is the zero-force limit of
$D_\parallel$ introduced above. For a lattice with
$q$ equally probable nearest-neighbor jumps, this
coefficient is given by
%%%%%%%%%%%%%%%%%%%%%%%%%%%%
\begin{equation}
D_0=\frac{1}{2q}\sum_{\Delta\mathbf r}
(\hat{\mathbf n}\cdot\Delta\mathbf r)^2.
\label{eq:D0}
\end{equation}
%%%%%%%%%%%%%%%%%%%%%%%%%%%%%
Substituting the Einstein relation into the leading-order
expression for the escape probability, we obtain
%%%%%%%%%%%%%%%%%%%%%%%%%%
\begin{equation}
1-Q_0\simeq
\frac{\Omega D_0|F_\parallel|}{a_\parallel}.
\label{eq:Q0_force}
\end{equation}
%%%%%%%%%%%%%%%%%%%%%%%%%
This result establishes the leading weak-force dependence
of the escape probability on the applied force and the
transverse cross-section. Together with the expression
for the mean longitudinal displacement derived in
Sec.~\ref{sec:Model}, it provides the basis for analyzing
confinement-induced mobility enhancement in the
following section.

\section{RESULTS}
\label{sec:Results}
The return probability obtained in Sec.~\ref{sec:ReturnProb},
together with the expression for the mean longitudinal
displacement in Eq.~\eqref{eq:mean_position}, allows us
to determine the response to an external force in a confined
channel. 
Our derivation applies to channels constructed
on any Bravais lattice, in arbitrary dimension, provided
the longitudinal direction and transverse periodic boundary
conditions are chosen as specified in Sec.~\ref{sec:ReturnProb}.
In particular, the two-dimensional square channel considered
in Ref.~\cite{Shafir2024} is a special case of this general
framework. 
As we show below, the resulting transport
exhibits a counterintuitive dependence on confinement:
in the glassy regime where $0<\alpha<1$, narrowing the channel
enhances the mean longitudinal displacement in the
weak-force regime. 
One might expect a wider channel to expedite transport by providing alternative paths around regions of strong trapping. 
But in the quenched trap model, increasing the transverse cross-section reduces the frequency of returns to previously visited sites. This exposes the particle to a larger number of distinct traps, producing the opposite effect.
In this section, we derive
the explicit dependence of the displacement on the force
and the transverse cross-section, evaluate it for simple
cubic and simple hexagonal lattices, and compare the
analytical predictions with numerical simulations.

To derive the explicit transport law, we start from
Eq.~\eqref{eq:mean_position}, where the return probability
$Q_0$ enters through the polylogarithm
$\mathrm{Li}_{-\alpha}(Q_0)$.
According to Eq.~\eqref{eq:Q0_weak}, the escape probability
satisfies $1-Q_0\simeq\Omega|v_\parallel|/a_\parallel$.
Thus, for a fixed finite transverse cross-section,
the weak-force limit corresponds to $Q_0\to1^-$.
Writing $\epsilon=1-Q_0$, we use the asymptotic relation
$\mathrm{Li}_{-\alpha}(1-\epsilon)
\sim\Gamma(1+\alpha)\epsilon^{-1-\alpha}$
as $\epsilon\to0^+$.
Substituting this relation into Eq.~\eqref{eq:mean_position},
and using $Q_0\to1$ in the numerator, we obtain
%%%%%%%%%%%%%%%%%%%%%%%%%%%%
\begin{equation}
\langle x_\parallel(t)\rangle
\sim
\frac{v_\parallel}{A\Gamma^2(1+\alpha)}
(1-Q_0)^{\alpha-1}t^\alpha.
\label{eq:mean_position_expanded}
\end{equation}
%%%%%%%%%%%%%%%%%%%%%%%%%%%%%%
Using Eq.~\eqref{eq:Q0_force} and the Einstein relation
$v_\parallel\simeq D_0F_\parallel$, we obtain,
for $F_\parallel>0$,
%%%%%%%%%%%%%%%%%%%%%%%%%%%%%
\begin{equation}
\langle x_\parallel(t)\rangle
\sim
\frac{(D_0F_\parallel)^\alpha}
{A\Gamma^2(1+\alpha)}
\left(\frac{a_\parallel}{\Omega}\right)^{1-\alpha}
t^\alpha.
\label{eq:universal_scaling}
\end{equation}
%%%%%%%%%%%%%%%%%%%%%%%%%%%%

For a fixed lattice, the geometric content of Eq.~\eqref{eq:universal_scaling} lies entirely in the
factor $(a_\parallel/\Omega)^{1-\alpha}$. In the glassy regime $0 < \alpha < 1$, the
exponent $1-\alpha$ is positive, so that under the same force and over the same
measurement time, a particle confined to a narrow channel advances further than when it is moving through a wider channel. 
Since $\Omega\geq 1$, the effect of $\Omega$ is bounded. The largest displacement occurs in the narrowest channel, $\Omega = 1$, where the walker advances along a single line of sites and transport becomes effectively one-dimensional. 
The prediction holds as long as the walk remains close to recurrence, that is, for $\Omega |v_\parallel| / a_\parallel \ll 1$, so that at fixed force the cross section cannot be made arbitrarily large.
%%%%%%%%%%%%%%%%%%%%%%%%%%%
\begin{figure}[t]
  \centering
  \includegraphics[width=\columnwidth]{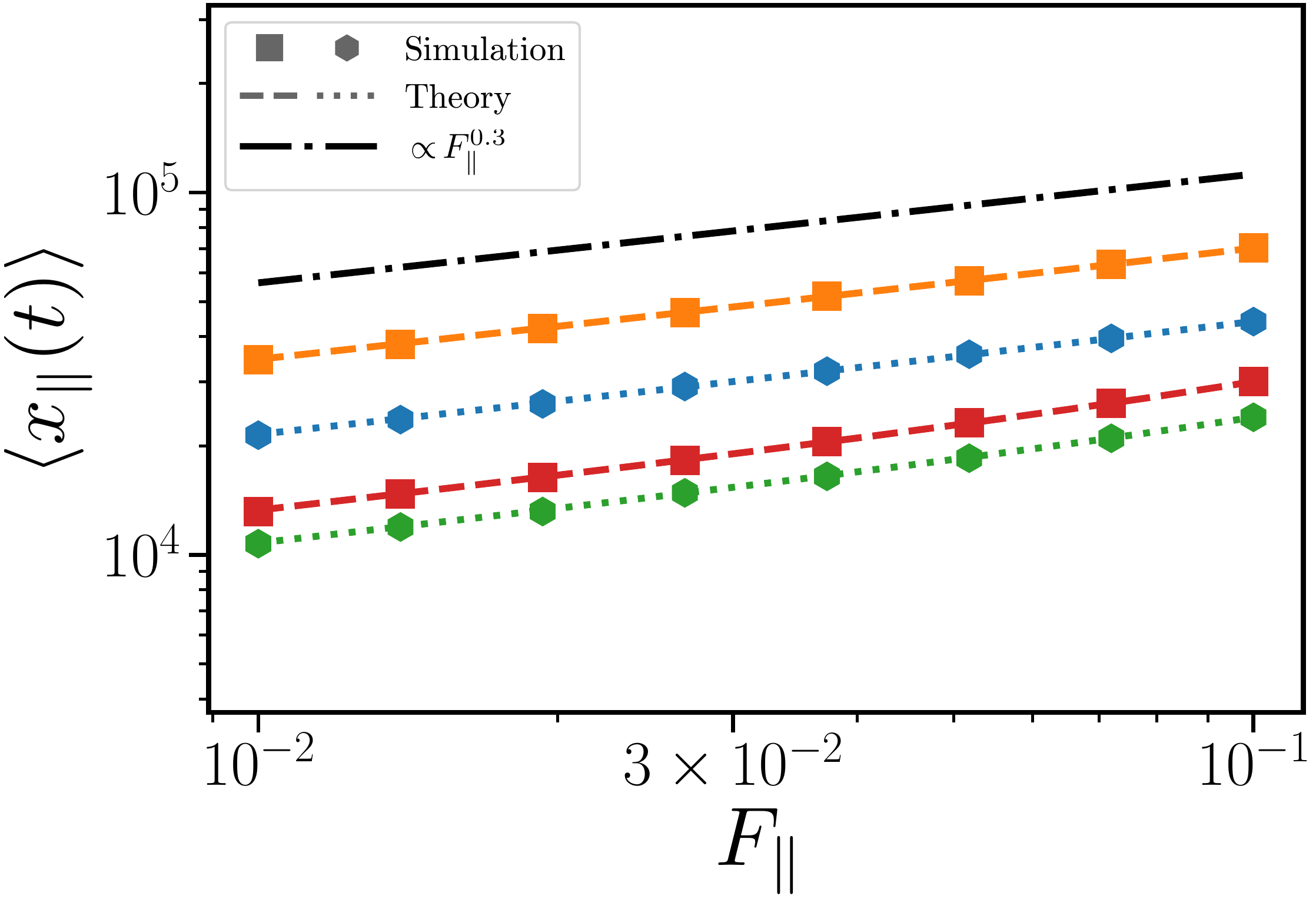}
  \caption{Mean longitudinal displacement $\langle x_\parallel(t)\rangle$
versus the applied force $F_\parallel$ for three-dimensional
channels of different transverse widths $w$.
Symbols (\legsimsq, \legsimhex) represent simulation results,
while dashed and dotted lines (\legthdash, \legthdot)
are the analytical predictions of Eq.~\eqref{eq:universal_scaling},
without fitting parameters.
Squares correspond to the simple cubic lattice, with
$w=2$~(\cubsym{mplorange}) and $w=4$~(\cubsym{mplred}),
while hexagons correspond to the simple hexagonal lattice,
with $w=2$~(\hexsym{mplblue}) and $w=3$~(\hexsym{mplgreen}).
The dash-dotted line indicates the predicted scaling
$F_\parallel^\alpha$.
Parameters: $R=10^5$, $t=10^{19}$, $\alpha=0.3$, and $A=1$.}
  \label{fig:force_dependence}
\end{figure}
%%%%%%%%%%%%%%%%%%%%%%%%%%%
To examine the predictions of Eq.~\eqref{eq:universal_scaling},
we consider two three-dimensional lattices: simple cubic
and simple hexagonal. The simple cubic lattice has six
nearest neighbors ($q=6$). For a channel oriented along
one of the cubic axes, the transverse cross-section is
square, and a width of $w$ lattice sites gives
$\Omega_{\mathrm{cub}}=w^2$.
The simple hexagonal lattice consists of triangular layers
stacked directly above one another, with six nearest
neighbors within each layer and two along the stacking
direction ($q=8$). For a channel oriented along the
stacking axis, the cross-section is a hexagonal patch of
a triangular layer. A patch containing $w$ sites along
each side has $\Omega_{\mathrm{hex}}=3w^2-3w+1$.
In both cases, the longitudinal lattice spacing is
$a_\parallel=a$.

To evaluate the longitudinal diffusion coefficient $D_0$,
we consider the diffusion tensor of the unbiased lattice
random walk, whose components are
$D_{ij}^{(0)}=(2q)^{-1}\sum_{\Delta\mathbf r}
\Delta r_i\Delta r_j$.
The coefficient $D_0$ introduced in Eq.~\eqref{eq:D0}
is the projection of this tensor onto the channel direction,
$D_0=\hat{\mathbf n}\cdot\mathbf D^{(0)}\hat{\mathbf n}$.
Choosing the $z$-axis along the channel, we first consider
the simple cubic lattice. Its six nearest-neighbor jumps
are directed along the positive and negative Cartesian
axes, giving the isotropic diffusion tensor
%%%%%%%%%%%%%%%%%%%%
\begin{equation}
   \mathbf D_{\mathrm{cub}}^{(0)}=
\left(
\begin{array}{ccc}
\dfrac{a^2}{6} & 0 & 0 \\[4pt]
0 & \dfrac{a^2}{6} & 0 \\[4pt]
0 & 0 & \dfrac{a^2}{6}
\end{array}
\right)
\end{equation}
%%%%%%%%%%%%%%%%%%%%%
Its projection onto the channel direction gives
%%%%%%%%%%%%%%%%%%%%
\begin{equation}
    \label{eq:D0cubic}
   D_{0,\mathrm{cub}}=\frac{a^2}{6}.
\end{equation}
%%%%%%%%%%%%%%%%%%%%
For the simple hexagonal lattice, the six nearest-neighbor
jumps within each triangular layer give identical diffusion coefficients in the
two transverse directions. The remaining two jumps
connect adjacent layers along the $z$-axis.
Taking the nearest-neighbor distance to be $a$ both
within and between layers, the unbiased diffusion
tensor is
%%%%%%%%%%%%%%%%%%%%
\begin{equation}
    \mathbf D_{\mathrm{hex}}^{(0)}
=
\left(
\begin{array}{ccc}
\dfrac{3a^2}{16} & 0 & 0 \\[4pt]
0 & \dfrac{3a^2}{16} & 0 \\[4pt]
0 & 0 & \dfrac{a^2}{8}
\end{array}
\right)
\end{equation}
%%%%%%%%%%%%%%%%%%%%%
Unlike the simple cubic lattice, the diffusion tensor
is anisotropic. Since the channel is oriented along
the stacking axis, its longitudinal diffusion coefficient
is
%%%%%%%%%%%%%%%%%%%%
\begin{equation}
    \label{eq:D0cubic}
    D_{0,\mathrm{hex}}=\frac{a^2}{8}.
\end{equation}
%%%%%%%%%%%%%%%%%%%%
For both lattices, the simple cubic and simple hexagonal, only the two jumps parallel and
antiparallel to the channel contribute to the
longitudinal diffusion coefficient. Thus,
Eq.~\eqref{eq:D0} gives $D_0=a^2/q$ for these geometries.

In Fig.~\ref{fig:force_dependence} we
present the behavior of the mean displacement as a function of the applied force for channels
of transverse width $w = 2$ to $w = 4$, in the case of simple cubic and simple hexagonal lattices. 
The simulations were performed according to the description of the model in Sec.~\ref{sec:Model} while the distribution of the quenched $\tau$s follows Eq.~\eqref{eq:waiting_time_pdf} for $\tau\geq\tau_0=\left[\frac{A}{\Gamma(1-\alpha)}\right]^{1/\alpha}$ and $0$ otherwise. 
The motion was recorded for $t=10^{19}$ with $A=1$ and $\alpha=0.3$, while averaging was performed over $R=10^5$ realizations. Periodic boundary conditions described in Sec.~\ref{sec:ReturnProb} were imposed. 
Excellent agreement between Eq.~\eqref{eq:universal_scaling} and the numerical results is observed. 
For both lattices and all widths considered, the mean displacement exhibits the predicted sublinear dependence $F_\parallel^{0.3}$.
The effect of the lattice structure and confinement appears in the amplitude of the displacement, determined by $\Omega$
and $D_0$.

\begin{figure}[t]
  \centering
  \includegraphics[width=\columnwidth]{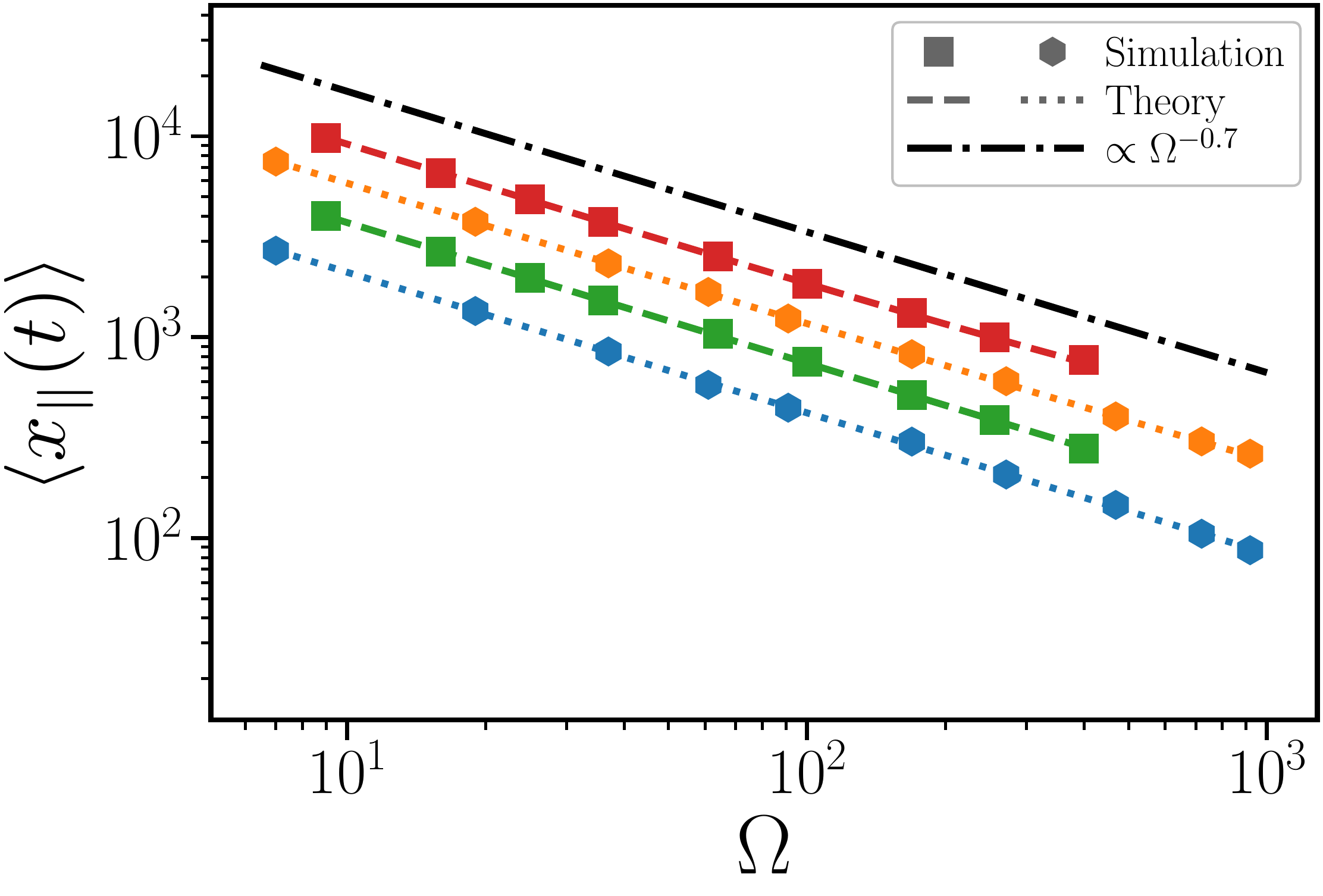}
  \caption{Mean longitudinal displacement $\langle x_\parallel(t)\rangle$
versus the number of sites in the transverse cross-section
$\Omega$ for three-dimensional channels under different
applied forces $F_\parallel$.
Symbols (\legsimsq, \legsimhex) represent simulation results,
while dashed and dotted lines (\legthdash, \legthdot)
are the analytical predictions of Eq.~\eqref{eq:universal_scaling},
without fitting parameters.
Squares correspond to the simple cubic lattice, with
$F_\parallel=10^{-3}$~(\cubsym{mplred}) and
$F_\parallel=5\times10^{-5}$~(\cubsym{mplgreen}),
while hexagons correspond to the simple hexagonal lattice,
with $F_\parallel=3\times10^{-4}$~(\hexsym{mplorange})
and $F_\parallel=10^{-5}$~(\hexsym{mplblue}).
The dash-dotted line indicates the predicted scaling
$\Omega^{-(1-\alpha)}$.
Parameters: $R=10^5$, $t=10^{19}$, $\alpha=0.3$, and $A=1$.}
  \label{fig:omega_dependence}
\end{figure}

In Fig.~\ref{fig:omega_dependence} we present the mean displacement as a
function of the number of sites in the cross-section $\Omega$ for four different values of the
force, two for each lattice, and again find excellent agreement with
Eq.~\eqref{eq:universal_scaling}. 
According to Eq.~\eqref{eq:universal_scaling} at fixed force the displacement scales as$\Omega^{-(1-\alpha)}$,
giving $\Omega^{-0.7}$ for $\alpha=0.3$. 
Namely, the motion in the 
wider channels is slower. 
This functional decay is independent of the lattice structure and set only by $\Omega$. 
The different decay amplitudes are determined by the combination
$(D_0 F_\parallel)^{\alpha}$ as shown in
Eq.~\eqref{eq:universal_scaling}.
%%%%%%%%%%%%%%%%%%%%%%
%%%%%%%%%%%%%%%%%%%%%%
%%%%%%%%%%%%%%%%%%%%
\begin{figure}[t]
  \centering
  \includegraphics[width=0.9\columnwidth]{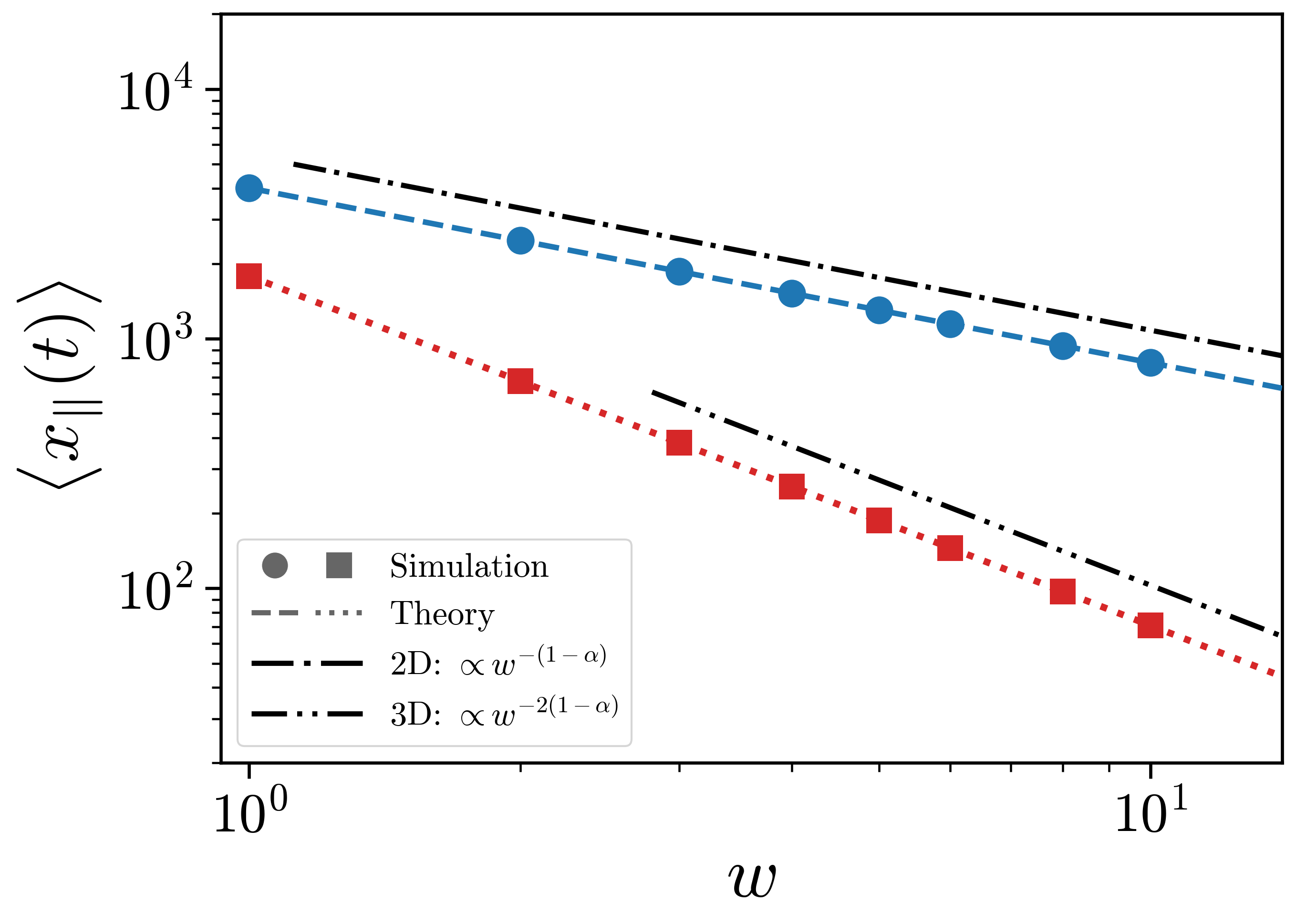}
  \caption{Mean longitudinal displacement $\langle x_\parallel(t)\rangle$
versus the transverse width $w$ for two anisotropic lattices.
Symbols (\legsimcirc, \legsimsq) represent simulation results,
while dashed and dotted lines (\legthdash, \legthdot)
are the analytical predictions of Eq.~\eqref{eq:universal_scaling},
without fitting parameters.
Circles correspond to a two-dimensional rectangular channel
($a=1$, $b=2$, $F_\parallel=2\times10^{-2}$)~(\rectsym{mplblue}),
while squares correspond to a three-dimensional orthorhombic
channel ($a=1$, $b=2$, $c=1.5$,
$F_\parallel=2\times10^{-3}$)~(\orthsym{mplred}).
The dash-dotted and dash-double-dotted lines indicate
the predicted scalings $w^{-(1-\alpha)}$ and
$w^{-2(1-\alpha)}$, respectively.
Parameters: $R=10^5$, $t=10^{14}$, $\alpha=0.3$, and $A=1$.}
  \label{fig:aniso}
\end{figure}
%%%%%%%%%%%%%%%%%%%%
The simple cubic and simple hexagonal lattices considered
above have a single lattice constant, so that the number
of sites in the transverse cross-section is directly related
to its physical size. To examine whether our results remain
valid for lattices with unequal spacings, we now consider
rectangular and orthorhombic lattices.
According to Eq.~\eqref{eq:universal_scaling}, the leading
dependence on transverse confinement is determined by
$\Omega$, while the longitudinal lattice spacing
$a_\parallel$ and the diffusion coefficient $D_0$
determine the remaining lattice-dependent factors.
In Fig.~\ref{fig:aniso}, we present the mean longitudinal
displacement as a function of the transverse width $w$
for a two-dimensional rectangular lattice with $a=1$
and $b=2$, where $\Omega=w$, and a three-dimensional
orthorhombic lattice with $a=1$, $b=2$, and $c=1.5$,
where $\Omega=w^2$.
The longitudinal direction is chosen along the lattice
vector of length $b$ in the rectangular case and $c$
in the orthorhombic case, so that $a_\parallel=b$ and
$a_\parallel=c$, respectively.
The numerical results agree with
Eq.~\eqref{eq:universal_scaling}, exhibiting the predicted
dependences $w^{-(1-\alpha)}$ and $w^{-2(1-\alpha)}$.
These results demonstrate that the leading confinement
scaling remains valid for lattices with unequal
longitudinal and transverse spacings.

%%%%%%%%%%%%%%%%%%
\begin{figure}[b]
  \centering
  \includegraphics[width=\columnwidth]{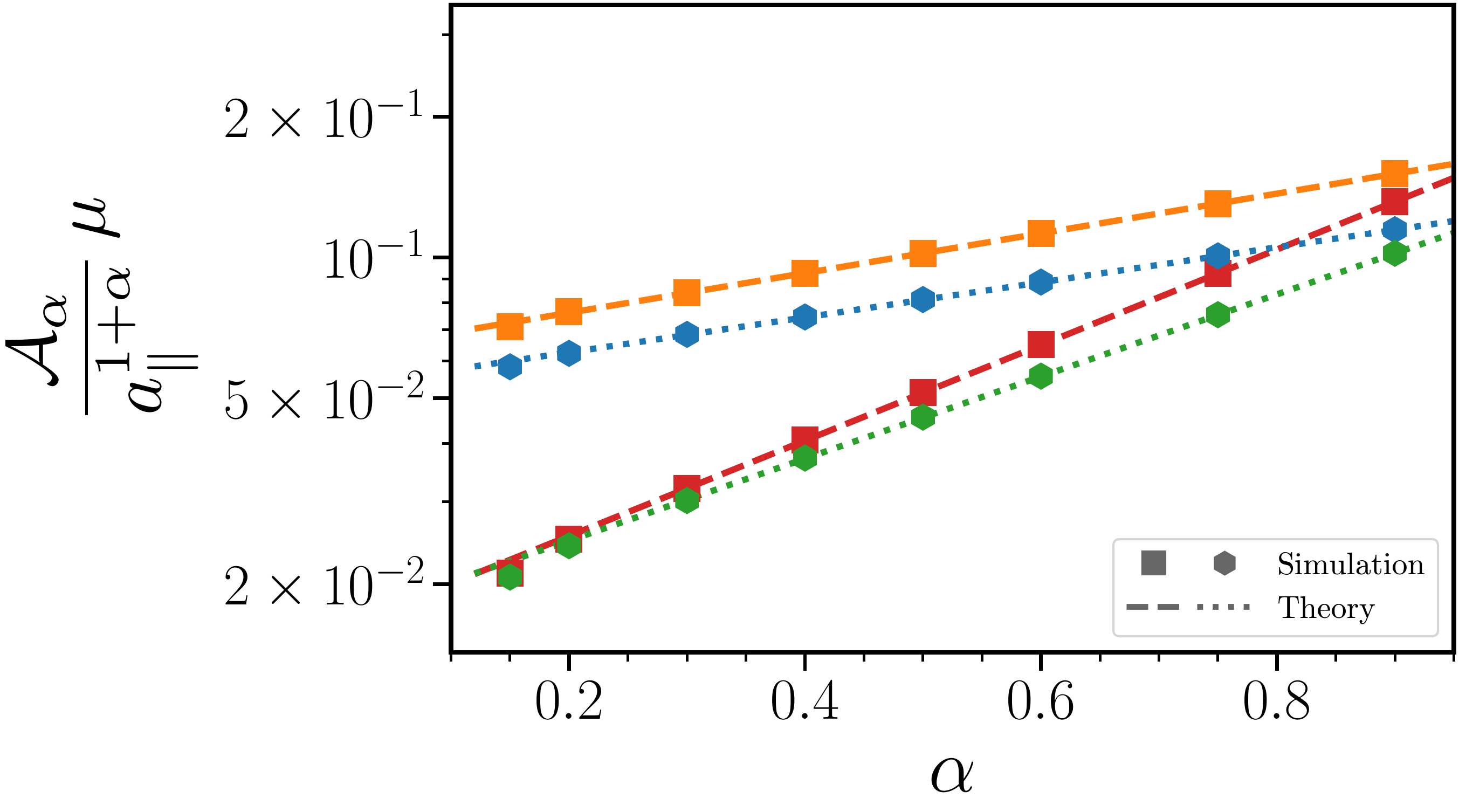}
  \caption{Normalized anomalous mobility
$\mathcal{A}_\alpha\mu/a_\parallel^{1+\alpha}$
as a function of the disorder exponent $\alpha$
for three-dimensional channels of different transverse widths $w$.
Symbols (\legsimsq, \legsimhex) represent simulation results,
while dashed and dotted lines (\legthdash, \legthdot)
are the analytical predictions obtained from
Eq.~\eqref{eq:log_relation}, without fitting parameters.
Squares correspond to the simple cubic lattice, with
$w=4$~(\cubsym{mplorange}) and $w=8$~(\cubsym{mplred}),
while hexagons correspond to the simple hexagonal lattice,
with $w=3$~(\hexsym{mplblue}) and
$w=5$~(\hexsym{mplgreen}).
Parameters: $R=10^5$, $t=10^{19}$,
$F_\parallel=10^{-3}$, and $A=1$.}
  \label{fig:alpha_dependence}
\end{figure}
%%%%%%%%%%%%%%%%%%

For ordinary transport, mobility is defined through a
linear relation between the mean displacement, the applied
force, and time. In the glassy regime considered here,
however, Eq.~\eqref{eq:universal_scaling} predicts a
nonlinear dependence on the applied force and sublinear
growth in time. To characterize this response, we use
the anomalous mobility $\mu$ via
$\langle x_\parallel(t)\rangle\sim\mu F_\parallel^\alpha
t^\alpha$, for $F_\parallel>0$.
Namely, Eq.~\eqref{eq:universal_scaling} yields
%%%%%%%%%%%%%%%%%%%%%%%%%%%%%%%%%%
\begin{equation}
\mu=
\frac{D_0^\alpha}{A\Gamma^2(1+\alpha)}
\left(\frac{a_\parallel}{\Omega}\right)^{1-\alpha}
\label{eq:anomalous_mobility}
\end{equation}
%%%%%%%%%%%%%%%%%%%
which holds for $0<\alpha<1$.
The mobility, therefore, in this anomalous phase of QTM depends not only on local disorder but also on the geometry of the media. 
Specifically, the factor $(a_\parallel/\Omega)^{1-\alpha}$ shows again the effect of diminished mobility in wider channels. 
The behavior of the mobility with $\alpha$ is nicely viewed as a function of $\alpha$ by defining $\mathcal{A}_\alpha \equiv A\,\Gamma^2(1+\alpha)$. 
Then 
%%%%%%%%%%%%%%%%%
\begin{equation}
    \ln \left[ \frac{\mathcal{A}_\alpha }{a_\parallel^{1+\alpha} } \mu\right]
    \simeq \alpha \ln \left( \frac{\Omega D_0}{a_\parallel^2} \right) - \ln \Omega .
    \label{eq:log_relation}
\end{equation}
%%%%%%%%%%%%%%%%%
Equation~\eqref{eq:log_relation} reveals how the relative
importance of transverse confinement and microscopic lattice
structure changes with $\alpha$.
In the limit $\alpha\to0^+$, its right-hand side approaches
$-\ln\Omega$, showing that the normalized mobility is
determined solely by the number of sites in the transverse
cross-section.
Conversely, as $\alpha\to1^-$, the expression approaches
$\ln(D_0/a_\parallel^2)$, which is independent of $\Omega$
and depends only on the microscopic lattice structure.
Thus, as $\alpha$ increases within the glassy regime,
the influence of transverse confinement gradually
diminishes relative to that of the underlying lattice
dynamics.
In Fig.~\ref{fig:alpha_dependence}, we compare the
predicted dependence on $\alpha$ with numerical simulations
for the simple cubic and simple hexagonal lattices.

\begin{figure}[t]
  \centering
  \includegraphics[width=\columnwidth]{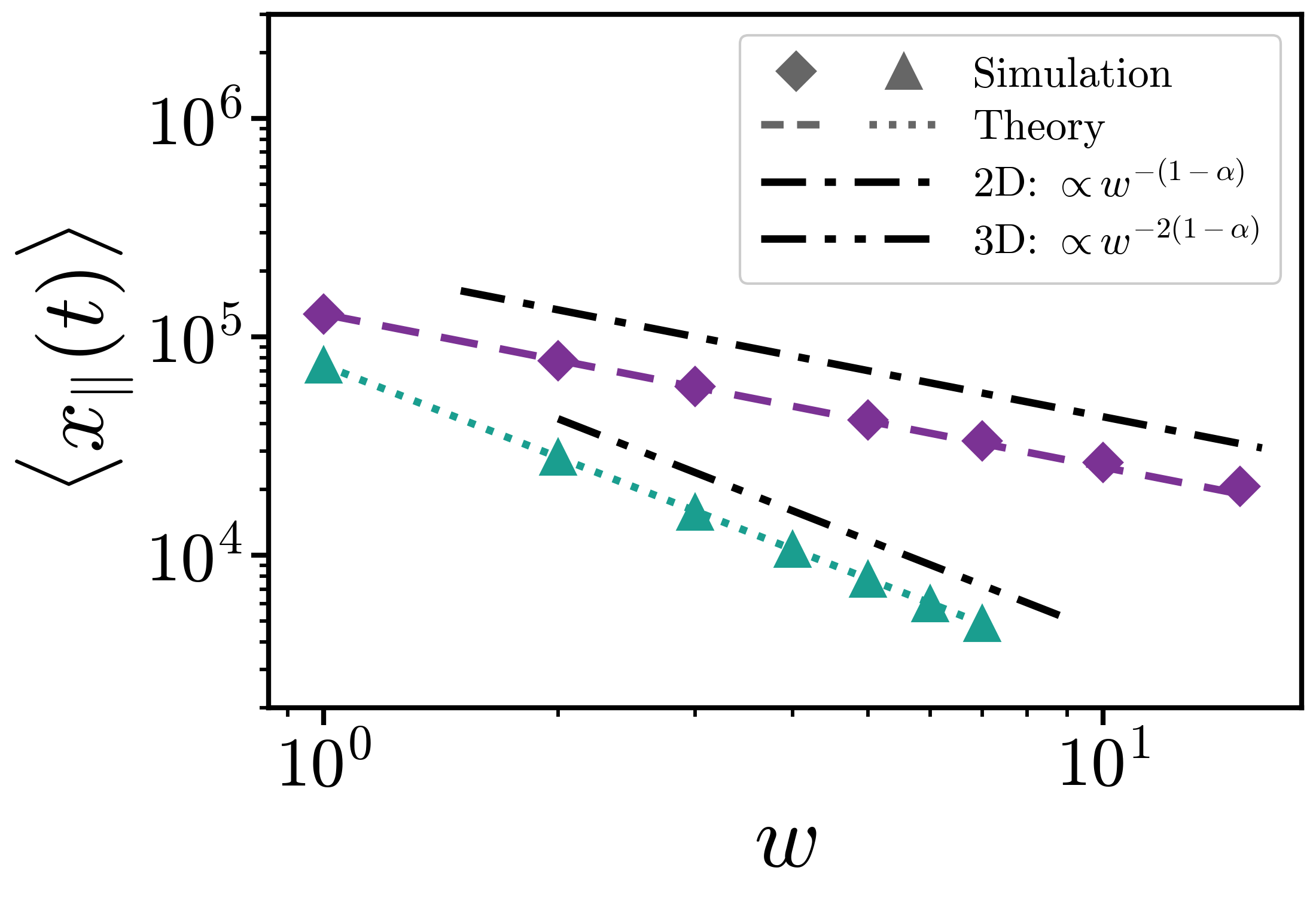}
  \caption{Mean longitudinal displacement $\langle x_\parallel(t)\rangle$
versus the transverse width $w$ for channels with reflecting
boundary conditions.
Symbols (\legsimdia, \legsimtri) represent simulation results,
while dashed and dotted lines (\legthdash, \legthdot)
are the analytical predictions of Eq.~\eqref{eq:universal_scaling},
derived for periodic boundary conditions, without fitting parameters.
Diamonds correspond to a two-dimensional square lattice,
with $F_\parallel=2\times10^{-2}$~(\diasym{rbcpurple}),
while triangles correspond to a three-dimensional simple cubic
lattice, with $F_\parallel=5\times10^{-3}$~(\trisym{rbcteal}).
A jump that would cross a wall is rejected, and the particle
remains at its current site for an additional dwell time
$\tau_{\mathbf r}$.
The dash-dotted and dash-double-dotted lines indicate
the predicted scalings $w^{-(1-\alpha)}$ and
$w^{-2(1-\alpha)}$, respectively.
Parameters: $a=1$, $R=10^5$, $t=10^{19}$,
$\alpha=0.3$, and $A=1$.}
  \label{fig:rbc}
\end{figure}

Although our analytical derivation assumes periodic transverse
boundary conditions, it is important to examine whether the
predicted mobility enhancement persists for reflecting walls.
We therefore performed additional simulations in which
attempted jumps across the channel boundaries are rejected,
and the particle remains at its current site for an additional
dwell time.
Unlike the periodic case, the resulting random walk is not
translationally invariant in the transverse directions.
This distinction is important because the mapping introduced
in Sec.~\ref{sec:Model}, particularly the expression for
$\Lambda$ in Eq.~\eqref{eq:lambda}, assumes equivalent
lattice sites and a site-independent return probability.
Furthermore, the calculation of $Q_0$ in
Sec.~\ref{sec:ReturnProb} explicitly relies on periodic
boundary conditions.
Nevertheless, as shown in Fig.~\ref{fig:rbc}, the numerical
results for two-dimensional square and three-dimensional
simple cubic channels agree with
Eq.~\eqref{eq:universal_scaling}, without fitting parameters.
The agreement extends beyond the predicted power-law
dependence on the transverse width. All the prefactors in the behavior of
the mean displacement seem to fit extremely well.
These results suggest that, at least for the geometries
and reflecting-boundary dynamics considered here,
the leading weak-force transport law is insensitive to
the choice between periodic and reflecting boundaries. The main and probably the most important ingredient is the confinement, as expressed by the presence of $\Omega$ in Eq.~\eqref{eq:universal_scaling}.
Establishing this behavior analytically remains an
open problem which we leave for future work. 

%%%%%%%%%%%%%%%%%%
%%%%%%%%%%%%%%%%%%
\section{Discussion}
\label{sec:Discussion}

In this work, we explored how confinement and geometric restrictions affect transport in the quenched trap model. 
It was recently discovered~\cite{Shafir2024} that in a 2-dimensional channel infinite in one direction and bounded in the other, the QTM on a square lattice exhibits intriguing behavior: the motion slows as the slab widens. 
Here we managed to generalize this result for a general channel in $d$ dimensions, unbounded along a single direction $\hat{\bf n}$. Our result holds for any Bravais lattice for the underlying random walk and periodic boundary conditions on the restricted geometries.  
The mathematical formula, obtained in the weak-driving limit, is elegant, as described by the mean displacement in Eq.~\eqref{eq:universal_scaling} and the anomalous mobility in Eq.~\eqref{eq:anomalous_mobility}. All macroscopic properties enter via the number of sites in the channel's transversal slice, $\Omega$. 
Surprisingly, this analytic form also describes the behavior when reflecting boundary conditions are applied. Although we don't have a mathematical proof, the numerical evidence supports this claim. 

We already mentioned that this phenomenon of slowing down in wider channels was also observed in completely different systems like molecules inside zeolites~\cite{brandani2024diffusion} and worm motion~\cite{prathyusha2026active}. 
The suggested general reason for this phenomenon, as we see it, is that widening the channel introduces additional configurations/states, and some of these new states can slow the motion. 
While this is a general qualitative suggestion, for QTM we can obtain a quantitative description of this scenario.  
Increasing the channel's transversal width increases the number of different sites $\Omega$ in the transverse slice. This in turn causes the random walk to sample more distinct sites, i.e., additional trapping times are sampled during the motion. 
The particle revisits fewer of the already sampled positions. 
Naively, it seems like a contradiction: if the particle is going back less frequently, it will probably move faster in the direction of the force. Indeed, the effect of $\Omega$ on the return probability, in the weak force limit, is exactly the same as the effect of the external drift $v_\parallel$ (see Eq.~\eqref{eq:Q0_weak}). 
The resolution to this controversy is in the statistics of the trapping times. 
Specifically, a broad power-law tail in $\psi(\tau)$ (see Eq.~\eqref{eq:waiting_time_pdf}). For $\alpha=T/T_g<1$, i.e., the glassy regime, the mean trapping time diverges. 
In this case, temporal condensation occurs~\cite{singh2023universal}, and the maximal trapping time is proportional to the total measurement time. 
This is the manifestation of the big jump principle~\cite{Vezzani2019,Vezzani2020LevyWalks,burov2026gaussian} that states: a large value of the sum (in our case the sum of the trapping times) is produced mainly by one unusually large summand rather than by the collective effect of many moderate summands. 
While the big jump principle usually affects rare events, for the broad power law considered in QTM it affects even the mean behavior, such as the particle's average position. 
This is why adding additional lattice sites, which lead to sampling of new trapping times by the random walk, slows transport. 
The motion is strongly affected by exceptionally long trapping times, and exploring more distinct sites increases the probability of encountering such a trap. 
Therefore, recurrence, usually associated with inefficient spatial exploration, can facilitate transport through a quenched trapping landscape.

%\vspace{-0.5cm}

\begin{acknowledgments}
This work was supported by the Israel Science Foundation, Grant No.~3791/25. 
The authors used ChatGPT (OpenAI) to assist with language editing, manuscript organization, and critical review of the clarity and consistency of scientific arguments.
\end{acknowledgments}

%\vspace{-0.3cm}
\appendix

\section{Evaluation of the Leading Zero-Mode Contribution}
\label{app:zero_mode}

In this appendix, we evaluate the zero-mode integral obtained
in Eq.~\eqref{eq:Wz_quadratic}. We write
\begin{equation}
\widetilde W_z^*(0)\simeq
\frac{a_\parallel}{2\pi\Omega}I_z,
\label{eq:app_Wz}
\end{equation}
where
\begin{equation}
I_z=\int_{-K}^{K}
\frac{dk_\parallel}
{\varepsilon+zD_\parallel k_\parallel^2
-izv_\parallel k_\parallel},
\label{eq:app_Iz}
\end{equation}
with $K=\pi/a_\parallel$ and $\varepsilon=1-z>0$.
We first take the limit $z\to1^-$ at fixed nonzero
longitudinal drift $v_\parallel$. In the main text we later consider
the weak-force limit $v_\parallel\to0$ for the obtained expression.
The order of these limits is important because the integrand
develops a singularity near $k_\parallel=0$ as
$\varepsilon\to0^+$.

To evaluate the integral, we separate the singular
contribution using the identity
\begin{equation}
\begin{aligned}
&\frac{1}{\varepsilon+zD_\parallel k_\parallel^2
-izv_\parallel k_\parallel}
\\
&\quad =
\frac{1}{\varepsilon-izv_\parallel k_\parallel}
\\
&\qquad -
\frac{zD_\parallel k_\parallel^2}
{(\varepsilon-izv_\parallel k_\parallel)
(\varepsilon+zD_\parallel k_\parallel^2
-izv_\parallel k_\parallel)}.
\end{aligned}
\label{eq:app_decomposition}
\end{equation}
We therefore write $I_z=I_z^{(1)}+I_z^{(2)}$,
where the first contribution is
\begin{equation}
I_z^{(1)}=
\int_{-K}^{K}
\frac{dk_\parallel}
{\varepsilon-izv_\parallel k_\parallel}.
\label{eq:app_I1}
\end{equation}
Multiplying the numerator and denominator by the
complex conjugate, we obtain
\begin{equation}
I_z^{(1)}=
\int_{-K}^{K}
\frac{\varepsilon+izv_\parallel k_\parallel}
{\varepsilon^2+z^2v_\parallel^2 k_\parallel^2}
\,dk_\parallel.
\label{eq:app_I1_conjugate}
\end{equation}
The imaginary part is odd in $k_\parallel$ and
therefore vanishes upon integration over the symmetric
interval. The remaining integral gives
\begin{equation}
I_z^{(1)}=
\frac{2}{z|v_\parallel|}
\arctan\left(
\frac{z|v_\parallel|K}{\varepsilon}
\right).
\label{eq:app_I1_arctan}
\end{equation}
Taking the limit $z\to1^-$, or equivalently
$\varepsilon\to0^+$, yields
\begin{equation}
\lim_{z\to1^-}I_z^{(1)}
=\frac{\pi}{|v_\parallel|}.
\label{eq:app_I1_limit}
\end{equation}

We now consider the second contribution,
\begin{equation}
\begin{aligned}
I_z^{(2)}=
-\int_{-K}^{K}
\frac{zD_\parallel k_\parallel^2\,dk_\parallel}
{(\varepsilon-izv_\parallel k_\parallel)
(\varepsilon+zD_\parallel k_\parallel^2
-izv_\parallel k_\parallel)}.
\end{aligned}
\label{eq:app_I2}
\end{equation}
In contrast to $I_z^{(1)}$, the singularity at
$k_\parallel=0$ is removed by the subtraction,
allowing us to take the limit $z\to1^-$ inside
the integral at fixed nonzero $v_\parallel$.
We obtain
\begin{equation}
\begin{aligned}
\lim_{z\to1^-}I_z^{(2)}
&=
\int_{-K}^{K}
\left[
\frac{D_\parallel}
{D_\parallel^2 k_\parallel^2+v_\parallel^2}
\right.\\
&\qquad\left.
-i\frac{D_\parallel^2 k_\parallel}
{v_\parallel(D_\parallel^2k_\parallel^2+v_\parallel^2)}
\right]dk_\parallel.
\end{aligned}
\label{eq:app_I2_limit}
\end{equation}
Again, the imaginary contribution is odd and integrates
to zero. Consequently,
\begin{equation}
\begin{aligned}
\lim_{z\to1^-}I_z^{(2)}
&=2\int_0^K
\frac{D_\parallel\,dk_\parallel}
{D_\parallel^2k_\parallel^2+v_\parallel^2}
\\
&=\frac{2}{|v_\parallel|}
\arctan\left(
\frac{D_\parallel K}{|v_\parallel|}
\right).
\end{aligned}
\label{eq:app_I2_arctan}
\end{equation}

Combining the two contributions and restoring
$K=\pi/a_\parallel$, we finally obtain
\begin{equation}
\widetilde W_1^*(0)\simeq
\frac{a_\parallel}{\pi\Omega|v_\parallel|}
\left[
\frac{\pi}{2}
+\arctan\left(
\frac{\pi D_\parallel}
{a_\parallel|v_\parallel|}
\right)
\right].
\label{eq:app_W1_arctan}
\end{equation}
This is the expression quoted in
Eq.~\eqref{eq:W1_arctan} of the main text.

\bibliography{enhancmentbib.bib}

@article{burov2026gaussian,
  title={A {G}aussian-Remainder Hierarchy for Sums of Random Variables with Big-Jump Statistics},
  author={Burov, Stanislav},
  journal={arXiv:2607.12357},
  year={2026}
}

@article{Vezzani2020LevyWalks,
  author  = {Vezzani, Alessandro and Barkai, Eli and Burioni, Raffaella},
  title   = {Rare events in generalized {L}\'{e}vy walks and the big jump principle},
  journal = {Scientific Reports},
  volume  = {10},
  pages   = {2732},
  year    = {2020}
}

@article{singh2023universal,
  title={Universal to nonuniversal transition of the statistics of rare events during the spread of random walks},
  author={Singh, RK and Burov, Stanislav},
  journal={Physical Review E},
  volume={108},
  number={5},
  pages={L052102},
  year={2023},
  publisher={APS}
}

@article{araujo2026sub,
  title={Sub-diffusion by random walks in quenched random environments},
  author={Ara{\'u}jo, Hugo A and Lanoisel{\'e}e, Yann and Pagnini, Gianni},
  journal={Physica A: Statistical Mechanics and its Applications},
  pages={131771},
  year={2026},
  publisher={Elsevier}
}

@article{shirataki2026observation,
  title={Observation-time-induced crossover in driven anomalous transport},
  author={Shirataki, Masahiro and Akimoto, Takuma},
  journal={Journal of Physics A: Mathematical and Theoretical},
  volume={59},
  number={22},
  pages={225001},
  year={2026},
  publisher={IOP Publishing}
}

@article{PhysRevE.69.026103,
  title = {Nonlinear response of the trap model in the aging regime: Exact results in the strong-disorder limit},
  author = {Monthus, C\'ecile},
  journal = {Phys. Rev. E},
  volume = {69},
  issue = {2},
  pages = {026103},
  numpages = {20},
  year = {2004},
  month = {Feb},
  publisher = {American Physical Society}
}

@article{PhysRevE.68.036114,
  title = {Anomalous diffusion, localization, aging, and subaging effects in trap models at very low temperature},
  author = {Monthus, C\'ecile},
  journal = {Phys. Rev. E},
  volume = {68},
  issue = {3},
  pages = {036114},
  numpages = {24},
  year = {2003},
  month = {Sep},
  publisher = {American Physical Society}
}

@article{aslangul1990dynamical,
  title={Dynamical exponents for one-dimensional random-random directed walks},
  author={Aslangul, Claude and Barthelemy, Marc and Pottier, No{\"e}lle and Saint-James, Daniel},
  journal={Journal of statistical physics},
  volume={59},
  number={1},
  pages={11--21},
  year={1990},
  publisher={Springer}
}

@article{derrida1983velocity,
  title={Velocity and diffusion constant of a periodic one-dimensional hopping model},
  author={Derrida, Bernard},
  journal={Journal of statistical physics},
  volume={31},
  number={3},
  pages={433--450},
  year={1983},
  publisher={Springer}
}

@article{RevModPhys.53.175,
  title = {Excitation dynamics in random one-dimensional systems},
  author = {Alexander, S. and Bernasconi, J. and Schneider, W. R. and Orbach, R.},
  journal = {Rev. Mod. Phys.},
  volume = {53},
  issue = {2},
  pages = {175--198},
  numpages = {0},
  year = {1981},
  month = {Apr},
  publisher = {American Physical Society}
}

@article{bernasconi1980diffusion,
  title={Diffusion and hopping conductivity in disordered one-dimensional lattice systems},
  author={Bernasconi, J and Schneider, WR and Wyss, W},
  journal={Zeitschrift f{\"u}r Physik B Condensed Matter},
  volume={37},
  number={2},
  pages={175--184},
  year={1980},
  publisher={Springer}
}

@article{bamb2025concentration,
  title={Concentration-dependent anomalous diffusion of crystal violet dye in agar gel: application of the continuous time random walk model},
  author={Bamb, Rachana D and Walimbe, Prasad C and Kulkarni, Sunil D and Kulkarni, Preeti S},
  journal={Physical Chemistry Chemical Physics},
  volume={27},
  number={12},
  pages={6212--6222},
  year={2025},
  publisher={The Royal Society of Chemistry}
}

@article{yuan2023hyperloop,
  title={Hyperloop-like diffusion of long-chain molecules under confinement},
  author={Yuan, Jiamin and Gao, Mingbin and Liu, Zhiqiang and Tang, Xiaomin and Tian, Yu and Ma, Gang and Ye, Mao and Zheng, Anmin},
  journal={Nature Communications},
  volume={14},
  number={1},
  pages={1735},
  year={2023},
  publisher={Nature Publishing Group UK London}
}

@article{brandani2024diffusion,
  title={Diffusion anomaly in nanopores as a rich field for theorists and a challenge for experimentalists},
  author={Brandani, Stefano and Hwang, Seungtaik and K{\"a}rger, J{\"o}rg and Mangano, Enzo},
  journal={Nature Communications},
  volume={15},
  number={1},
  pages={5721},
  year={2024},
  publisher={Nature Publishing Group UK London}
}

@article{qi2026confinement,
  title={Confinement-Driven Anomalous Behaviors for Diffusion in Zeolites: Mechanisms and Beyond},
  author={Qi, Ji and Yuan, Jiamin and Liu, Zhiqiang and Zheng, Anmin},
  journal={Accounts of Chemical Research},
  volume={59},
  number={13},
  pages={2044--2055},
  year={2026},
  publisher={ACS Publications}
}

@article{yang2025nonmonotonic,
  title={Nonmonotonic enhancement of nanorod diffusion during 1D to quasi-2D confinement transition},
  author={Yang, Jingbin and Zhang, Jia and Shen, Tianshun and Dong, Ruo-Yu},
  journal={Colloids and Surfaces A: Physicochemical and Engineering Aspects},
  volume={724},
  pages={137453},
  year={2025},
  publisher={Elsevier}
}

@article{bandyopadhyay1995diffusion,
  title={Diffusion anomaly in silicalite and VPI-5 from molecular dynamics simulations},
  author={Bandyopadhyay, Sanjoy and Yashonath, Subramanian},
  journal={The Journal of Physical Chemistry},
  volume={99},
  number={12},
  pages={4286--4292},
  year={1995},
  publisher={ACS Publications}
}

@article{prathyusha2026active,
  title={Active polymers translocate faster in confinement},
  author={Prathyusha, KR and Sarkar, Paulami and Xu, Justin and Bhamla, Saad},
  journal={Physical Review Letters},
  volume={137},
  number={5},
  pages={058302},
  year={2026},
  publisher={APS}
}

@book{Abramowitz1988,
  title={Handbook of Mathematical Functions with Formulas, Graphs, and Mathematical Tables},
  author={Abramowitz, M. and Stegun, I. A. and Romer, R. H.},
  publisher={American Association of Physics Teachers},
  year={1988}
}

@article{Akimoto2018,
  title={Non-self-averaging behaviors and ergodicity in quenched trap models with finite system sizes},
  author={Akimoto, Takuma and Barkai, Eli and Saito, Keiji},
  journal={Phys. Rev. E},
  volume={97},
  pages={052143},
  year={2018},
  publisher={APS}
}

@article{Akimoto2020,
  title={Trace of anomalous diffusion in a biased quenched trap model},
  author={Akimoto, T. and Saito, K.},
  journal={Phys. Rev. E},
  volume={101},
  pages={042133},
  year={2020}
}

@article{Arous2015,
  title={Randomly trapped random walks},
  author={Ben Arous, G. and Cabezas, M. and \v{C}ern\'{y}, J. and Royfman, R.},
  journal={Ann. Probab.},
  volume={43},
  pages={2405},
  year={2015}
}

@article{Barkai2001,
  title={Fractional Fokker-Planck equation, solution, and application},
  author={Barkai, E.},
  journal={Phys. Rev. E},
  volume={63},
  pages={046118},
  year={2001},
  publisher={APS}
}

@article{Benichou2011,
  title={Intermittent search strategies},
  author={B{\'e}nichou, Olivier and Loverdo, Claude and Moreau, Michel and Oshanin, Gleb},
  journal={Reviews of Modern Physics},
  volume={83},
  number={1},
  pages={81},
  year={2011},
  publisher={APS}
}

@article{Benichou2018,
  title={Tracer diffusion in crowded narrow channels},
  author={B\'{e}nichou, O. and Illien, P. and Oshanin, G. and Sarracino, A. and Voituriez, R.},
  journal={J. Phys. Condens. Matter},
  volume={30},
  pages={443001},
  year={2018}
}

@article{Bouchaud1990,
  title={Anomalous diffusion in disordered media: Statistical mechanisms, models and physical applications},
  author={Bouchaud, Jean-Philippe and Georges, Antoine},
  journal={Physics Reports},
  volume={195},
  number={4-5},
  pages={127--293},
  year={1990},
  publisher={Elsevier}
}

@article{Bouchaud1992,
  title={Weak ergodicity breaking and aging in disordered systems},
  author={Bouchaud, Jean-Philippe},
  journal={Journal de Physique I},
  volume={2},
  number={9},
  pages={1705--1713},
  year={1992},
  publisher={EDP Sciences}
}

@article{Burov2007,
  title={Occupation time statistics in the quenched trap model},
  author={Burov, S. and Barkai, E.},
  journal={Phys. Rev. Lett.},
  volume={98},
  pages={250601},
  year={2007},
  publisher={APS}
}

@article{Burov2011,
  title={Time transformation for random walks in the quenched trap model},
  author={Burov, S. and Barkai, E.},
  journal={Phys. Rev. Lett.},
  volume={106},
  pages={140602},
  year={2011},
  publisher={APS}
}

@article{Burov2012,
  title={Weak subordination breaking for the quenched trap model},
  author={Burov, S. and Barkai, E.},
  journal={Phys. Rev. E},
  volume={86},
  pages={041137},
  year={2012}
}

@article{Burov2017,
  title={From quenched disorder to continuous time random walk},
  author={Burov, Stanislav},
  journal={Physical Review E},
  volume={96},
  number={5},
  pages={050103},
  year={2017},
  publisher={APS}
}

@article{Burov2020,
  title={The transient case of the quenched trap model},
  author={Burov, Stanislav},
  journal={Journal of Statistical Mechanics: Theory and Experiment},
  volume={2020},
  number={7},
  pages={073207},
  year={2020},
  publisher={IOP Publishing}
}

@article{Cerny2015,
  title={Randomly trapped random walks on {$\mathbb{Z}^d$}},
  author={\v{C}ern\'{y}, J. and Wassmer, T.},
  journal={Stoch. Proc. Appl.},
  volume={125},
  pages={1032},
  year={2015}
}

@book{Feller1968,
  title={An Introduction to Probability Theory and Its Applications, Vol. 1},
  author={Feller, William},
  volume={1},
  year={1968},
  publisher={John Wiley \& Sons}
}

@book{Gradshteyn2014,
  title={Table of Integrals, Series, and Products},
  author={Gradshteyn, I. S. and Ryzhik, I. M.},
  edition={8th},
  year={2014},
  publisher={Academic Press},
  address={Burlington}
}

@book{Hughes1995,
  title={Random Walks and Random Environments: Volume 1: Random Walks},
  author={Hughes, Barry D.},
  year={1995},
  publisher={Clarendon Press},
  address={Oxford}
}

@book{Klafter2011,
  title={First Steps in Random Walks: From Tools to Applications},
  author={Klafter, J. and Sokolov, I. M.},
  publisher={Oxford University Press},
  address={Oxford},
  year={2011}
}

@article{Kubo1966,
  title={The fluctuation-dissipation theorem},
  author={Kubo, Ryogo},
  journal={Reports on Progress in Physics},
  volume={29},
  number={1},
  pages={255},
  year={1966},
  publisher={IOP Publishing}
}

@article{Machta1985,
  title={Random walks on site disordered lattices},
  author={Machta, J.},
  journal={J. Phys. A},
  volume={18},
  pages={L531},
  year={1985}
}

@article{Malgaretti2019,
  title={Special issue on transport in narrow channels},
  author={Malgaretti, P. and Oshanin, G. and Talbot, J.},
  journal={J. Phys. Condens. Matter},
  volume={31},
  pages={270201},
  year={2019}
}

@article{Metzler2000,
  title={The random walk's guide to anomalous diffusion: a fractional dynamics approach},
  author={Metzler, Ralf and Klafter, Joseph},
  journal={Physics Reports},
  volume={339},
  number={1},
  pages={1--77},
  year={2000},
  publisher={Elsevier}
}

@article{Monthus1996,
  title={Models of traps and glass phenomenology},
  author={Monthus, C{\'e}cile and Bouchaud, Jean-Philippe},
  journal={Journal of Physics A: Mathematical and General},
  volume={29},
  number={14},
  pages={3847},
  year={1996},
  publisher={IOP Publishing}
}

@article{Monthus2003,
  title={Anomalous diffusion, localization, aging, and subaging effects in trap models at very low temperature},
  author={Monthus, C{\'e}cile},
  journal={Phys. Rev. E},
  volume={68},
  pages={036114},
  year={2003},
  publisher={APS}
}

@article{Montroll1965,
  title={Random Walks on Lattices. {II}},
  author={Montroll, Elliott W. and Weiss, George H.},
  journal={J. Math. Phys.},
  volume={6},
  number={2},
  pages={167--181},
  year={1965},
  publisher={AIP Publishing}
}

@book{Redner2001,
  title={A Guide to First-Passage Processes},
  author={Redner, Sidney},
  year={2001},
  publisher={Cambridge University Press},
  address={Cambridge}
}

@article{Shafir2022,
  title={The case of the biased quenched trap model in two dimensions with diverging mean dwell times},
  author={Shafir, Dan and Burov, Stanislav},
  journal={Journal of Statistical Mechanics: Theory and Experiment},
  volume={2022},
  number={3},
  pages={033301},
  year={2022},
  publisher={IOP Publishing}
}

@article{Shafir2024,
  title={Disorder-Induced Anomalous Mobility Enhancement in Confined Geometries},
  author={Shafir, Dan and Burov, Stanislav},
  journal={Phys. Rev. Lett.},
  volume={133},
  pages={037101},
  year={2024},
  publisher={APS}
}

@article{Vezzani2019,
  title={Single-big-jump principle in physical modeling},
  author={Vezzani, Alessandro and Barkai, Eli and Burioni, Raffaella},
  journal={Phys. Rev. E},
  volume={100},
  pages={012108},
  year={2019},
  publisher={APS}
}

@book{Weiss1994,
  title={Aspects and Applications of the Random Walk},
  author={Weiss, George H.},
  year={1994},
  publisher={North-Holland},
  address={Amsterdam}
}

@article{Bertin2003response,
  author  = {Bertin, E. M. and Bouchaud, J.-P.},
  title   = {Linear and nonlinear response in the aging regime of the one-dimensional trap model},
  journal = {Phys. Rev. E},
  volume  = {67},
  pages   = {065105},
  year    = {2003}
}

@article{Barkai1998,
  author  = {Barkai, E. and Fleurov, V. N.},
  title   = {Generalized Einstein relation: A stochastic modeling approach},
  journal = {Phys. Rev. E},
  volume  = {58},
  pages   = {1296},
  year    = {1998}
}
\end{document}